\documentclass{osa-article}
\journal{osajournal}
\articletype{Research Article}
\usepackage{mathtools}
\usepackage{braket}
\usepackage{graphicx}
\usepackage{xcolor}
\usepackage{enumitem}

\DeclarePairedDelimiterX{\norm}[1]{\lVert}{\rVert}{#1}
\newcommand{\parashort}[1]{\noindent \textbf{#1}}
\newcommand{\imodel}{\mathbf{I}}
\newcommand{\fmodel}{\mathbf{F}}

\begin{document}
\title{SPDCinv: Inverse Quantum-Optical Design of High-Dimensional Qudits}

\author{Eyal Rozenberg,\authormark{1,*} Aviv Karnieli,\authormark{2} Ofir Yesharim,\authormark{3} Joshua Foley-Comer,\authormark{3} Sivan Trajtenberg-Mills,\authormark{2,4} Daniel Freedman,\authormark{5} Alex M. Bronstein,\authormark{1} and Ady Arie\authormark{3}}

\address{\authormark{1}Department of Computer Science, Technion, Haifa, Israel\\
\authormark{2}Raymond and Beverly Sackler School of Physics and Astronomy, Tel Aviv University, Israel\\
\authormark{3}School of Electrical Engineering, Fleischman Faculty of Engineering, Tel Aviv University, Israel\\
\authormark{4}Massachusetts Institute of Technology, Cambridge, MA, USA \\
\authormark{5}Google Research, Haifa, Israel \\}
\email{\authormark{*}eyalr@campus.technion.ac.il}

\begin{abstract}  \label{abstract}
Spontaneous parametric down-conversion (SPDC) in quantum optics is an invaluable resource for the realization of high-dimensional qudits with spatial modes of light. One of the main open challenges is how to directly generate a desirable qudit state in the SPDC process. This problem can be addressed through advanced computational learning methods; however, due to difficulties in modeling the SPDC process by a fully differentiable algorithm that takes into account all interaction effects, progress has been limited. Here, we overcome these limitations and introduce a physically-constrained and differentiable model, \textcolor{black}{validated against experimental results for shaped pump beams and structured crystals,} capable of learning every interaction parameter in the process. We avoid any restrictions induced by the stochastic nature of our physical model and integrate the dynamic equations governing the evolution under the SPDC Hamiltonian. We solve the inverse problem of designing a nonlinear quantum optical system that achieves the desired quantum state of down-converted photon pairs. The desired states are defined using either the second-order correlations between different spatial modes or by specifying the required density matrix. By learning nonlinear volume holograms as well as different pump shapes, we successfully show how to generate maximally entangled states. Furthermore, we \textcolor{black}{simulate} all-optical coherent control over the generated quantum state by actively changing the profile of the pump beam. Our work can be useful for applications such as novel designs of high-dimensional quantum key distribution and quantum information processing protocols. In addition, our method can be readily applied for controlling other degrees of freedom of light in the SPDC process, such as the spectral and temporal properties, and may even be used in condensed-matter systems having a similar interaction Hamiltonian.
\end{abstract}

\section{Introduction} \label{sec:introduction}
The penetration of advanced machine learning (ML) methods into physics has led to far-reaching advances in both theoretical predictions and experiments, yielding exciting and interesting new results \cite{carleo2017solving, raissi2019physics, iten2020discovering, choo2020fermionic, gentile2021learning, karniadakis2021physics}. Some of the most interesting progress has come from the solution of inverse problems \cite{tarantola2005inverse} aimed at finding novel experimental setups that produce a desired physical observable \cite{krenn2016automated, melnikov2018active, tamayo2018automatic, malkiel2018plasmonic, molesky2018inverse, yao2019intelligent, minkov2020inverse, jagtap2020conservative, krenn2020computer, colburn2021inverse, wiecha2021deep}. Nevertheless, there are still physical phenomena, particularly in quantum physics, that have yet benefit from this progress. This may be attributed at least partially to the lack of appropriate computational tools for modelling complex quantum systems, and in some cases to the stochastic dynamics involved in modelling quantum phenomena such as spontaneous processes and fluctuations of quantum fields \cite{sinatra2002truncated, brambilla2004simultaneous, corney2015non, lewis2016approximate, drummond2017higher, weinbub2018recent, trajtenberg2020simulating}. 

One important branch of quantum physics that might benefit significantly from the adoption of inverse design algorithms is quantum optics \cite{scully1999quantum, garrison2008quantum}. Quantum optics has proven to be an invaluable resource for the realization of many quantum technologies, such as quantum communication \cite{ursin2007entanglement,gisin2007quantum,vallone2015experimental,chen2021integrated}, quantum computing \cite{knill2001scheme,kok2007linear,spring2013boson,zhong2020quantum}, and cryptography \cite{bennett1992experimental,bennett2020quantum,sit2017high,liao2017satellite,pirandola2020advances}. A prominent reason for this is the availability of sources for generating nonclassical light \cite{garrison2008quantum}, which are mainly based on nonlinear interactions \cite{boyd2020nonlinear}. The most prevalent of these processes is spontaneous parametric down-conversion (SPDC) in second order nonlinear $\chi^{(2)}$ materials \cite{SPDCreview2018}. The nonlinear coefficient of ferroelectric materials can be modulated by electric field poling in two out of the three crystal axes \cite{berger1998nonlinear,broderick2000hexagonally}. Recently, this capability was extended to enable modulation in all three axes using focused laser beams  \cite{xu2018three,wei2018experimental,liu2019nonlinear,wei2019efficient,imbrock2020waveguide,liu2020nonlinear,zhang2021nonlinear, chen2021quasi, arie2021storing}. The 3D nonlinear photonic crystals (NLPCs) offer a promising new avenue for shaping and controlling arbitrary quantum correlations between photons. This new technology introduces additional degrees of freedom for tailoring the quantum state of structured photon-pairs \cite{walborn2012generalized,malik2016multi,dosseva2016shaping,kovlakov2017spatial,erhard2018twisted,PhysRevA.98.060301,cui2019wave, erhard2020advances, boucher2021engineering}. Solving the inverse quantum optical design would make it possible to find the optimal physical parameters of the system, such as the pump beam profile and the nonlinear 2D and 3D volume holograms embedded in the NLPC, that yield the desired quantum state. These capabilities can be used for the generation of maximally entangled photonic states of arbitrary dimensionality that allow stronger violation of generalized Bell’s inequalities, the encoding of larger capacities of quantum information on light \cite{brandt2020high}, and improved security in quantum key distribution \cite{krenn2015twisted,sit2017high,sit2018quantum}.

If we wish to employ ML methods for problems in quantum optics, it is crucial to have a good physical model of the quantum optical process in question and integrate it into the algorithm itself \cite{raissi2019physics, de2019deep, PhysRevLett.124.010508, jagtap2020conservative, pang2020physics, SIRIGNANO2020109811, karniadakis2021physics, pakravan2021solving}. The model should ideally encompass the relevant conservation laws, physical principles, and phenomenological behaviors. Such physically-constrained models will ensure convergence to physically realizable solutions, reduce the parameter search, improve the predictive accuracy and statistical efficiency of the model, and allow for faster training with improved generalization. However, there are obstacles to incorporating ML into quantum optics while still properly capturing the physics. In order to account for general optical medium geometry, diffraction, dispersion, and non-perturbative effects in non-classical light generation (such as SPDC), accurate simulation schemes must be employed, that go beyond the scope of the more frequently used analytic calculations \cite{torres2003quantum, walborn2012generalized, SPDCreview2018,kolobov1999spatial}.  However, such models – which are more appealing for the inverse design of complex optical media – are often stochastic \cite{kolobov1999spatial, PhysRevA.69.023802, trajtenberg2020simulating}. The stochastic nature of the problem, also prominent in other physical fields such as those which employ Monte Carlo simulations \cite{binder1993monte},  makes modern descent-based algorithms difficult to employ.

\begin{figure}[ht]
  \centering
  \includegraphics[width=1\textwidth]{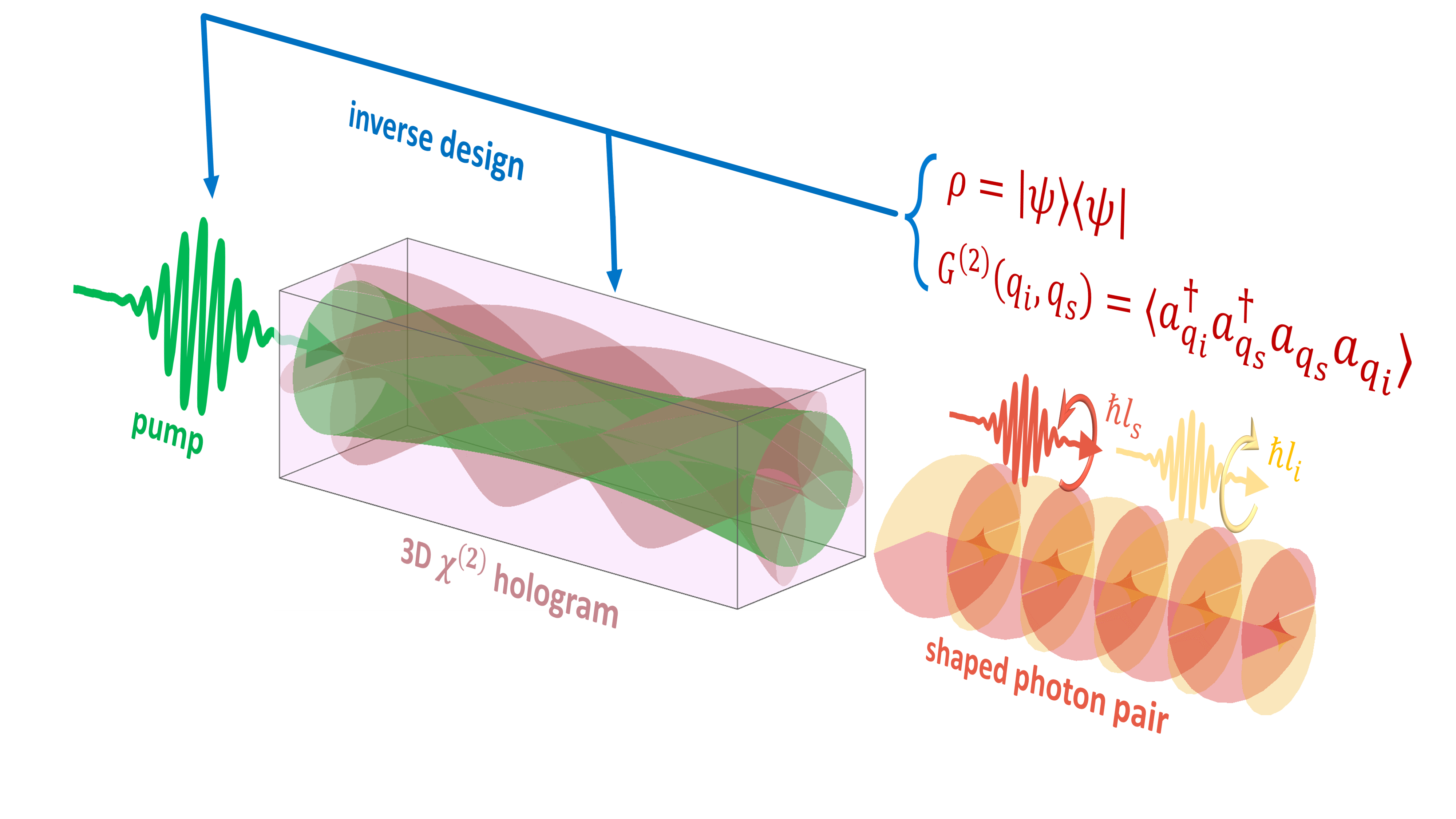}
  \caption{ \small An illustration of the inverse design problem: Given the desired coincidence rate counts, $G^{(2)}$, and density matrix, $\rho$, the \emph{SPDCinv} algorithm solves the inverse design problem and extracts the optimal quantum volume hologram, embedded in 3D NLPC, and the complex pump beam structure, for generating the desired quantum state of the spontaneously emitted structured photon-pairs.}
  \label{fig:illustration}
\end{figure}

In this paper, we solve the inverse design problem of generating structured and entangled photon pairs in quantum optics using tailored nonlinear interactions in the SPDC process. The learned interaction parameters can then be used to predict the generation of the desired quantum state or correlations between structured photon-pairs in future experiments, as illustrated in Fig. \ref{fig:illustration}. Our \emph{SPDCinv} model captures the full dynamics (the governing dynamics derived from Heisenberg's equations of motion), takes into account high-order interaction effects, and can learn every parameter of the quantum optical process. We show how to make an inherent stochastic description of SPDC fully differentiable, making it amenable to descent based methods of optimization. Furthermore, we use a split-step Fourier (SSF) method \cite{stoffa1990split} to solve our forward model. To the best of our knowledge, this is the first time that a differentiable model has been integrated with SSF -- a feature which is also relevant for many other inverse problems in optics and quantum mechanics (it combines diffraction, or more generally, propagation in space, to solve nonlinear partial differential equations, like the nonlinear Schr\"{o}dinger equation). \textcolor{black}{Our forward model has already been validated against a number of published experimental results, detailed in Refs. \cite{trajtenberg2020simulating, DiDomenico:21, DiDomenico:Talk}, for the cases of structured pump beams \cite{kovlakov2017spatial, PhysRevA.98.060301, mair2001entanglement} and structured crystals \cite{trajtenberg2020simulating, DiDomenico:21, DiDomenico:Talk}. In this paper, we further validate it against other experiments \cite{kovlakov2017spatial,PhysRevA.69.023802}, obtaining very good agreement for both on-axis spatial mode correlations, as well as to the quantum state tomography of the generated state. Moreover, we demonstrate the full process of inverse design to obtain the correct relations between crystal length and pump waist, as achieved in the experiments \cite{kovlakov2017spatial}.}

We use our model to discover the optimal quantum volume holograms (embedded in 2D \cite{berger1998nonlinear, broderick2000hexagonally, chowdhury2001experimental, ellenbogen2009nonlinear,bloch2012twisting,shapira2012two,hong2014nonlinear,zhu2020high} or 3D NLPCs \cite{xu2018three,wei2018experimental,liu2019nonlinear,wei2019efficient,imbrock2020waveguide,liu2020nonlinear,zhang2021nonlinear, chen2021quasi, arie2021storing}) and the pump structures that generate desired nontrivial quantum correlations (coincidence rate counts) and quantum states (bi-photon density matrices). We demonstrate the generation of high-dimensional maximally entangled photon-pairs and show how the generated quantum state and its correlations can be controlled entirely optically using shaped pump fields interacting with the initially-learned 3D NLPC hologram -- a feature that can find applications in qudit-based quantum key distribution and quantum information protocols that work at high switching rates. Our \emph{SPDCinv} model has been made available at \cite{jax-spdc_inv}.
\footnote{A preliminary short abstract of this work was presented at the CLEO 2021 conference \cite{rozenberg2021inverse}.}.

\section{Algorithmic Design} \label{sec:algorithm}

\subsection{Methodology}
The procedure for the study of inverse problems in physical systems can be divided into the following three steps \cite{tarantola2005inverse, aster2018parameter}: (i) identifying a minimal set of model parameters whose values completely characterize the system; (ii) identifying the physical laws and dynamics governing the system; and (iii) the use of actual results to infer the values of the model parameters. Given a desired observable-set, $\mathbb{O}_d$, describing the quantum state or any related features, our goal is to find the unknown physical parameters, $\Lambda$, that characterize the system,

\begin{equation}
\label{eq:inverse_model}
\Lambda = \imodel(\mathbb{O}_d),
\end{equation}
where $\imodel(\cdot)$ is our inverse solver. We physically constrain our \emph{SPDCinv} model by integrating it with the interaction dynamics of the SPDC process. In this manner, the model captures the interaction properties, such as diffraction, space-dependent nonlinear coupling, vacuum fluctuations and non-perturbative effects. The dynamics of SPDC is prescribed by the Heisenberg equations of motion: $i\hbar \partial_t \hat{E} = [\hat{E},\hat{H}_{\mathrm{SPDC}}]$, for the field operators $\hat{E}$ evolving under the SPDC Hamiltonian $\hat{H}_{\mathrm{SPDC}}$, where $\hbar$ is the reduced Planck's constant. To solve the dynamics, it is enough to consider two pairs of c-number coupled wave equations along the 3D interaction medium, in terms of the field operator matrix elements \cite{trajtenberg2020simulating}, given as: 
\begin{equation}
\begin{split}
    i\frac{\partial E_{i}^{out}}{\partial \zeta}  = -\frac{\nabla^2_\perp}{2k_i} E_{i}^{out}+\kappa_ie^{-i\Delta k \zeta}(E_{s}^{vac})^* \\
    i\frac{\partial E_{i}^{vac}}{\partial \zeta}  =   -\frac{\nabla^2_\perp}{2k_i}E_{i}^{vac}+\kappa_ie^{-i\Delta k \zeta}(E_{s}^{out})^* \\
    i\frac{\partial E_{s}^{out}}{\partial \zeta}  = -\frac{\nabla^2_\perp}{2k_s}E_{s}^{out}+\kappa_se^{-i\Delta k \zeta}(E_{i}^{vac})^*\\
    i\frac{\partial E_{s}^{vac}}{\partial \zeta}  = -\frac{\nabla^2_\perp}{2k_s}E_{s}^{vac}+\kappa_se^{-i\Delta k \zeta}(E_{i}^{out})^*
\label{eq:waveeq}
\end{split}
\end{equation}
where $\zeta=z$ is the coordinate along the direction of propagation. In the above equation: $E_{j}^{out},E_{j}^{vac}$ ($j=i,s$ for the idler and signal fields respectively) are the output and vacuum field amplitudes of the generated photons and vacuum fluctuations; $\nabla^2_\perp$ is the transverse Laplacian operator; $k_j$ is the wavenumber; $\kappa_{j} (\textbf{r}, \zeta)=\frac{\omega_j^2}{c^2 k_j} \chi^{(2)} (\textbf{r}, \zeta) \mathcal{E}_{p}(\mathbf{r})$ is the nonlinear-coupling coefficient, where $\textbf{r}=(x,y)$ is a position on the transverse plane; $\chi^{(2)}(\textbf{r}, \zeta)$ stands for the (spatially varying) second-order susceptibility and $\mathcal{E}_{p}(\mathbf{r})$ is the (spatially varying) pump field envelope; $c$ is the speed of light in vacuum; and $\Delta k=k_p-k_s-k_i$ is the phase mismatch. The quantum vacuum noise is emulated by initializing a large number of instances of Gaussian noise in both the idler and signal fields (denoted as $E_i^{vac}$ and $E_s^{vac}$ in Eq. \ref{eq:waveeq}), creating the physical vacuum field uncertainty.  We summarize Eq. \ref{eq:waveeq} in a compact fashion by denoting all of the fields as $E = (E_{i}^{out}, E_{i}^{vac}, E_{s}^{out}, E_{s}^{vac})$, and writing
\begin{equation}
    i\frac{\partial E}{\partial \zeta} = \mathcal{L}(\Lambda)E
    \label{eq:waveeq_compact}
\end{equation}
where $\mathcal{L}$ is the operator given by the righthand side of Eq. \ref{eq:waveeq}, and $\Lambda$ represents the list of physical parameters described in the previous exposition.  In practice, we will be particularly interested in the pump field $\mathcal{E}_{p}$ and second-order susceptibility $\chi^{(2)}$, that is, $\Lambda = (\mathcal{E}_{p}(\cdot), \chi^{(2)}(\cdot))$, with all other parameters being taken as fixed.  However, we note that the formulation which follows is general, and does not depend on the parameters of interest. \textcolor{black}{We emphasize that our model is not semiclassical, but instead fully equivalent to the solution of the quantum Heisenberg equations of motion for the field operators \cite{brambilla2004simultaneous, trajtenberg2020simulating}. In our case, we assume the signal and idler fields are initially in the vacuum state (which was rigorously shown to justify the random sampling of vacuum fluctuations \cite{brambilla2004simultaneous, trajtenberg2020simulating}), and employ the fact that the generated multimode squeezed vacuum state belongs to the family of Gaussian states (see Eq. \ref{eq:G2}). Further, we note that similar approaches, for example ones that sample the Wigner function at random to calculate quantum observables, are also employed in quantum optics and condensed matter theory \cite{sinatra2002truncated, corney2015non, lewis2016approximate, drummond2017higher, weinbub2018recent}}.

We integrate the fields along the direction of propagation according to Eq. \ref{eq:waveeq}, and solve the coupled wave equations for the large ensemble of quantum vacuum realizations in parallel. We use a time-unfolded version \cite{gregor2010learning} of the SSF method \cite{stoffa1990split, agrawal2001applications} to solve for the propagation along the crystal. Then, we derive the second-order statistics to describe the resulting quantum state; an approach that was validated against experimental results, for several cases of shaped pump beams and structured crystals \cite{PhysRevA.69.023802, kovlakov2017spatial, trajtenberg2020simulating, DiDomenico:21, DiDomenico:Talk} \textcolor{black}{(see also section \ref{subsec:validation}).} This strategy facilitates differentiation back through the model and enables application of the latest optimization methods for learning its physical parameters, thereby overcoming issues related to the fundamentally stochastic nature of the model.

In what follows, we shall refer to the solution of Eq. \ref{eq:waveeq} (or alternatively Eq. \ref{eq:waveeq_compact}), together with the mapping onto a particular set of observables of interest, denoted as $\mathbb{O}$, as our \textit{forward model}. In particular, we write 
\begin{equation}
\label{eq:forward_model}
\mathbb{O} = \fmodel(\Lambda).
\end{equation}
Given a desired observable set, $\mathbb{O}_d$ , the general inverse problem involves finding the physical parameters $\Lambda$ which produce it. We specialize by solving a parameterized version of the inverse problem.  In particular, suppose that the physical parameters of interest $\Lambda$ depend upon parameters $\theta$ which specify them, i.e. $\Lambda = \Lambda(\theta)$.  Such parameters $\theta$ may, for example, be coefficients of basis expansions; we will see concrete examples shortly. In this case, we solve the inverse problem by solving the optimization problem
\begin{equation}
\label{eq:optimizer}
\theta^* = \min_\theta \mathcal{D}\big( \fmodel(\Lambda(\theta)), \mathbb{O}_d \big)
\end{equation}
In the above, $\mathcal{D}(\cdot, \cdot)$ is a discrepancy measure between two sets of observables.  For example, we may take $\mathcal{D}(\mathbb{O}, \mathbb{O}') = \|\mathbb{O} - \mathbb{O}' \|_\beta$, where $\|\cdot\|_\beta$ is the Euclidean $\beta$-norm; alternatively, if the observables are normalized to unit $1$-norm, then $\mathcal{D}$ can be the Kullback-Leibler divergence.  In the case where we are measuring the discrepancy between two density matrices, we may take $\mathcal{D}$ to be the Trace Distance \cite{rana2016trace}. In Eq. \ref{eq:optimizer}, we are therefore trying to minimize the discrepancy between the set of observables given by a particular parameter specification $\theta$, and the desired set of observables $\mathbb{O}_d$.  The inverse model is then given by
\begin{equation}
\label{eq:model}
\imodel(\mathbb{O}_d) = \Lambda(\theta^*)
\end{equation}

In order to solve the optimization problem in Eq. \ref{eq:optimizer}, an approach based on gradient descent may be employed.  The key is that the forward model of Eq. \ref{eq:waveeq}, while quite complicated, can be expressed in such a way that it is fully differentiable.  As a result, any library which can auto-differentiate a system may be used to compute the relevant gradients, thereby allowing for the solution to the optimization problem in Eq. \ref{eq:optimizer}.  In practice, we use JAX \cite{jax2018github}, a Python library designed for high-performance numerical computing and automatic differentiation.

\begin{figure}[h]
  \centering
  \includegraphics[width=0.95\textwidth]{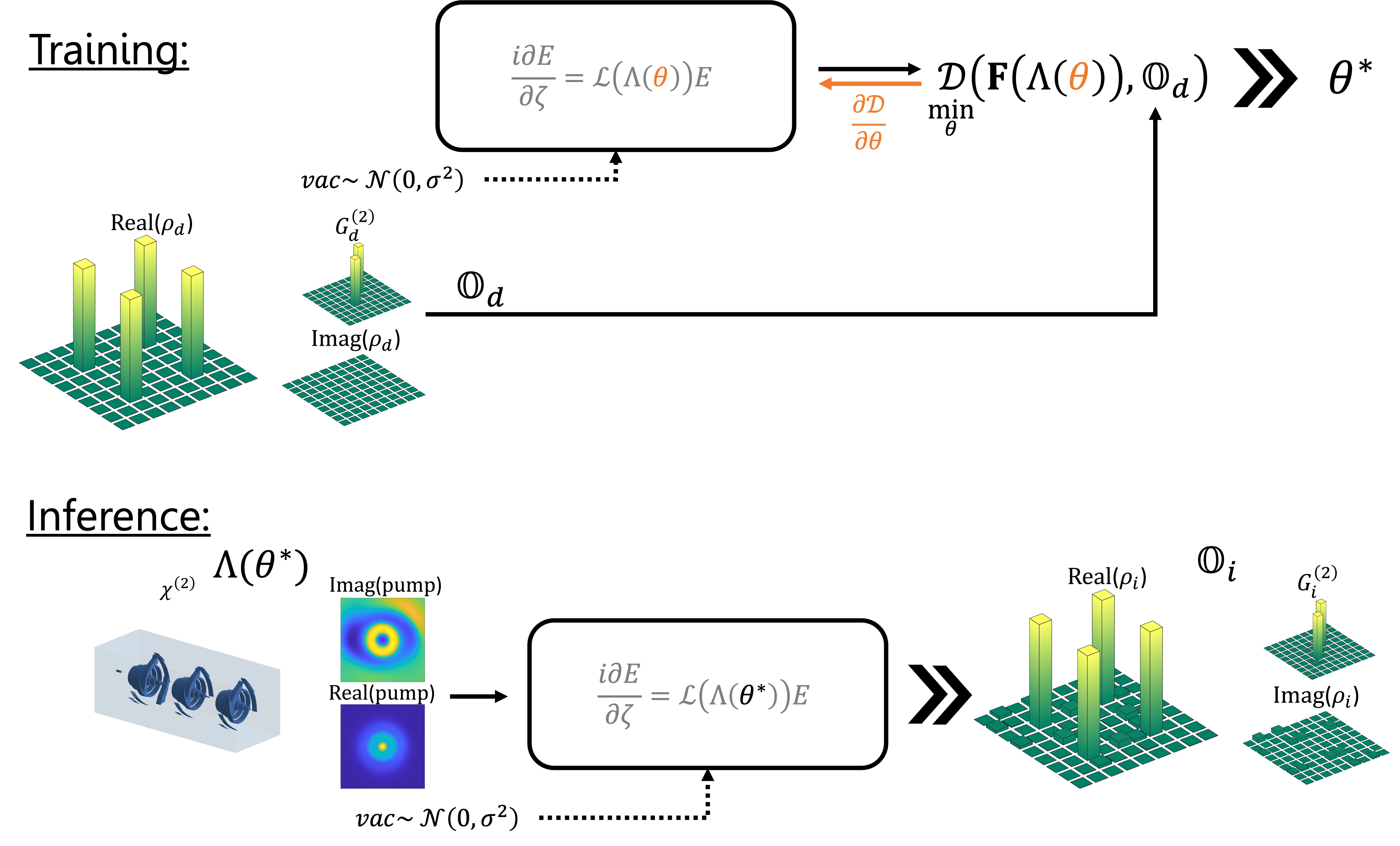}
  \caption{ \small Description of the \emph{SPDCinv} algorithm in two phases. (1) In the training phase (upper panel), the parameterized version of the inverse design is solved. The model receives as input the desired observables and emits the parameterization of the physical parameters that will produce it, by solving the optimization problem. The learning process is described by applying gradient descent (in orange) to the appropriate discrepancy measure, $\mathcal{D}(\cdot, \cdot)$. (2) In the inference phase (lower panel), the model receives the computed physical parameters and emits the observables. The compact notation of the partial differential equation refers to the solution of the Heisenberg equations, Eq. \ref{eq:waveeq}. The quantum vacuum noise is integrated externally (dashed line).}
  \label{fig:model_paradigm}
  \vspace{-0.5cm}
\end{figure}

Finally, given the solution to the inverse problem, we may run the forward model to compute the observables that actually result from the interaction parameters we have computed, that is
\begin{equation}
    \label{eq:inference}
    \mathbb{O}_i = \fmodel(\Lambda(\theta^*))
\end{equation}
where the subscript $i$ indicates \emph{inference}.  The degree to which the inferred observables $\mathbb{O}_i$ match the desired observables $\mathbb{O}_d$ will indicate the quality of the inverse algorithm.  The overall algorithm is summarized in Fig. \ref{fig:model_paradigm}.

\parashort{Interaction Parameters:} We may learn any physical parameters $\Lambda$ of the interaction, e.g. wavelength, temperature profile, poling period, poling profile, etc. In this work, the 2D/3D NLPC structure, $\chi^{(2)}(\textbf{r}, \zeta)$, and pump beam profile, $\mathcal{E}_{p}(\textbf{r})$, are the unknown physical parameters we seek to learn, that is $\Lambda = (\mathcal{E}_{p}(\cdot), \chi^{(2)}(\cdot))$. We parameterize the 2D/3D crystal hologram and pump beam profile by the multi-dimensional parameters $\theta_\mathcal{E}$ and $\theta_\chi$, respectively, such that $\Lambda(\theta) = (\mathcal{E}_{p}(\cdot; \theta_\mathcal{E}), \chi^{(2)}(\cdot; \theta_\chi))$.  We now discuss in more detail how this parameterization is performed.

The parameters we learn can be as general as we want, subject to technological and physical restrictions. To decrease the dimensionality of learned parameters in order to ensure smoother convergence of the inverse problem's solution, the continuous functions of the NLPC holograms are represented using a finite set of unknowns. One way to do this is through expansion in set basis functions that are mutually orthogonal, which may also change as a function of $\zeta$; the parameters $\theta$ then include the coefficients of the expansion. Examples include the Hermite-Gauss (HG) and Laguerre-Gauss (LG) bases, though many other possibilities exist. These basis functions are often scaled according to a transverse length, which for light beams is usually referred to as the waist size, a term which we adopt hereafter for all basis functions. Learning the waist sizes of each of the basis functions individually adds further degrees of freedom to our model. The exact role of the parameters can be seen by formally writing the NLPC structure and the pump profile as a linear combination of the basis functions:
\begin{align}
    \chi^{(2)}(\textbf{r}, \zeta; \theta_\chi) & = \sum_{n=1}^{N_\chi}{\alpha_\chi^n \Phi_\chi^n(\textbf{r}, \zeta; w_\chi^n)} & \quad \theta_\chi = \left\{ \left(\alpha_\chi^n, w_\chi^n \right) \right\}_{n=1}^{N_\chi} \notag \\
    \mathcal{E}_{p}(\textbf{r}; \theta_\mathcal{E}) & = \sum_{n=1}^{N_\mathcal{E}}{\alpha_\mathcal{E}^n \Phi_\mathcal{E}^n(\textbf{r}; w_\mathcal{E}^n)} & \quad \theta_\mathcal{E} = \left\{ \left(\alpha_\mathcal{E}^n, w_\mathcal{E}^n \right) \right\}_{n=1}^{N_\mathcal{E}}
\label{eq:parameterized}
\end{align}
where $\alpha_\chi^n, \alpha_\mathcal{E}^n$ are the learned basis coefficients; $w_\chi^n, w_\mathcal{E}^n$ are the learned basis function waist sizes; and $\Phi_\chi^n, \Phi_\mathcal{E}^n$ are the basis functions. Here, the basis function index $n$ sums over both transverse modal numbers, for example the orbital angular momentum $l$- and radial $p$-indices for LG modes.

\subsection{Observables}\label{subsec:observables}
The set of desired observables describing the generated quantum state is given by the coincidence rate count, $G^{(2)}$, and density matrix of the bi-photon quantum state, $\rho$, such that in general $\mathbb{O}_d = (G^{(2)}_d, {\rho}_d)$. Their evaluation is achieved by first solving the Heisenberg equations of motion for the SPDC Hamiltonian over a large number of independent realizations of the vacuum noise, projecting the output and noise fields onto a desired orthonormal basis of optical modes, and then taking the ensemble average to obtain first-order correlations \cite{PhysRevA.69.023802, trajtenberg2020simulating}, which (for the signal) is given by $G^{(1)}(q_s,q'_s)=\braket{\psi|a^{\dagger}_{q_s}a_{q'_s}|\psi}$. Here, $\ket{\psi}$ denotes the quantum state, $a$ ($a^{\dagger}$) denotes the photon annihilation (creation) operator, and $q_s$ denotes any quantum number of the signal photon, for example, LG modes, HG modes, etc. Second-order correlations are derived using the fact that the quantum state of SPDC, the squeezed vacuum state \cite{wu1986generation}, belongs to the family of Gaussian states, for which all higher-order correlations can be obtained from the first-order ones \cite{gardiner2004quantum}. The coincidence rate is given by the second-order quantum correlation function, which determines the probability of finding an idler photon in mode $q_i$ and a signal photon in mode $q_s$

\begin{equation}
G^{(2)}(q_i,q_s,q_s,q_i)=\braket{\psi|a^{\dagger}_{q_i}a^{\dagger}_{q_s}a_{q_s}a_{q_i}|\psi}
\label{eq:G2}
\end{equation}

To extract the optimal model parameters that generate the desired quantum correlations over a given basis, we solve the optimization problem in Eq. \ref{eq:optimizer}. Here, $\mathcal{D}(\cdot, \cdot)$ is taken as a typical measure of discrepancy between two probability distributions. For example, we may use the Kullback-Leibler divergence \cite{georgiou2003kullback}, the L1 norm \cite{gine2003bm}, or an ensemble of both.

To obtain the full quantum state generated by the SPDC process, we use quantum state tomography (QST) \cite{thew2002qudit, agnew2011tomography, toninelli2019concepts}. Eq. \ref{eq:G2} allows for the calculation of any coincidence measurement performed on the system, on any basis of our choice. Since the process of QST involves a sequence of projective coincidence measurements on different bases, we can readily reconstruct  the density matrix, $\rho$, of the entangled two-qudit state, through a series of linear operations. Here, naturally, $\mathcal{D}(\cdot, \cdot)$ (in Eq. \ref{eq:optimizer}) is taken to be the Trace Distance \cite{rana2016trace} -- a metric on the space of density matrices that measures the distinguishability between two states.

The tomographic reconstruction is performed using the correlation data collected from the projections of the simulated bi-photon state onto orthogonal as well as mutually unbiased bases (MUBs) \cite{toninelli2019concepts, agnew2011tomography}. The density matrix of the bi-photon system can be written as
\begin{equation}
    \rho  = \frac{1}{d^{2}}
    \sum_{m,n=0}^{d^{2}-1}\rho_{mn}\\
    \sigma_{m}\otimes\sigma_{n}
\end{equation}
where $\sigma_{m}$ are the set of generators that span the $d$-dimensional tomography space (for example, Pauli and Gell-Mann matrices for $d=2$ and $3$, respectively). The expansion coefficients $\rho_{mn}$ are found via 
\begin{equation}
    \rho_{mn}  =
    \sum_{i,j=0}^{d-1}\\
    a_{m}^{i}a_{n}^{j}\braket{\lambda_{m}^{i}\lambda_{n}^{j}|\\
    \rho|\lambda_{m}^{i}\lambda_{n}^{j}}
\end{equation}
with $a_{m}^{i}$ and $|\lambda_{m}^{i}\rangle$ denoting the $i^{th}$ eigenvalue and eigenstate of $\sigma_{m}$, respectively \cite{toninelli2019concepts}. The required projections inside the sum function are found in a similar manner to Eq. \ref{eq:G2}, with the pure basis states replaced by the MUBs, when necessary.

\section{Results} \label{sec:results}

The proposed method can be readily employed to generate  desired quantum correlations between SPDC structured photon-pairs. Further, by emulating QST integrated into the learning stage, we can tailor specific, high-dimensional quantum states desirable for photonic quantum information and communication. In this section, we use our algorithm to solve the inverse design problem and extract the optimal quantum volume holograms, embedded in 2D or 3D NLPCs, and the complex pump beam structures for generating desired second-order quantum correlations or density matrices. We either let our algorithm learn the NLPC volume holograms, the complex pump beam profiles, or both. We discover that the quantum state of SPDC photons and their correlations can be all-optically controlled, by first learning the crystal volume holograms with a given pump mode, and then changing the initial pump mode in inference phase. This active optical control has the advantage of altering the quantum state in a non-trivial manner, while retaining its purity. Further, we find that learning the quantum volume hologram and the pump beam profile simultaneously can improve the accuracy of the generated results, in comparison with the desired state. The \emph{SPDCinv} training phase takes about one hour on 4 nvidia t4 16gb gpus, for all configurations involving 1mm-long NLPCs.

\subsection{Model Validation}\label{subsec:validation}
\begin{figure*}[]
\centering
\begin{tabular}{c}
    \includegraphics[width=0.95\linewidth]{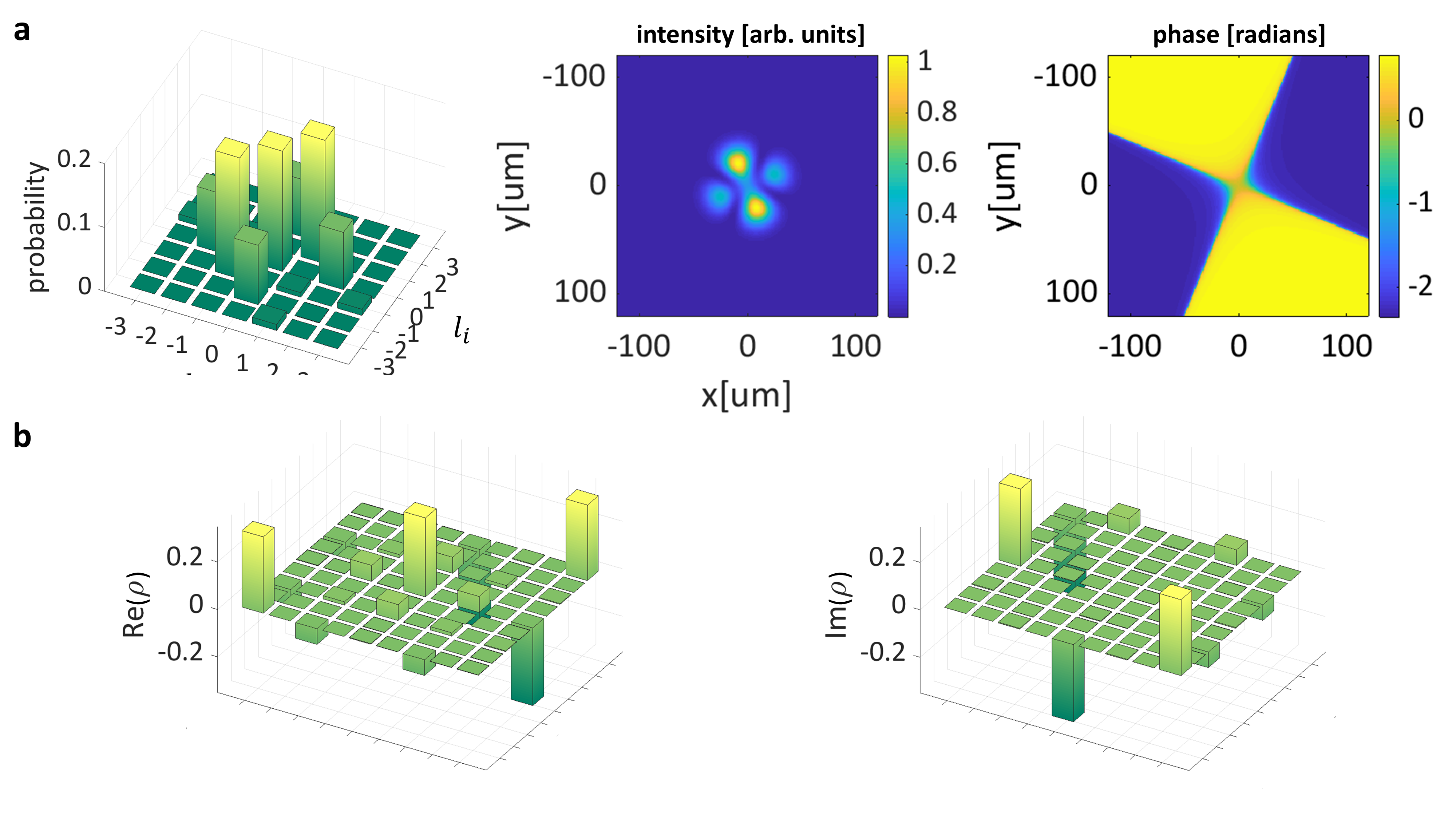}  \\
    
    \includegraphics[trim={0 9cm 0 0},clip,width=0.95\linewidth]{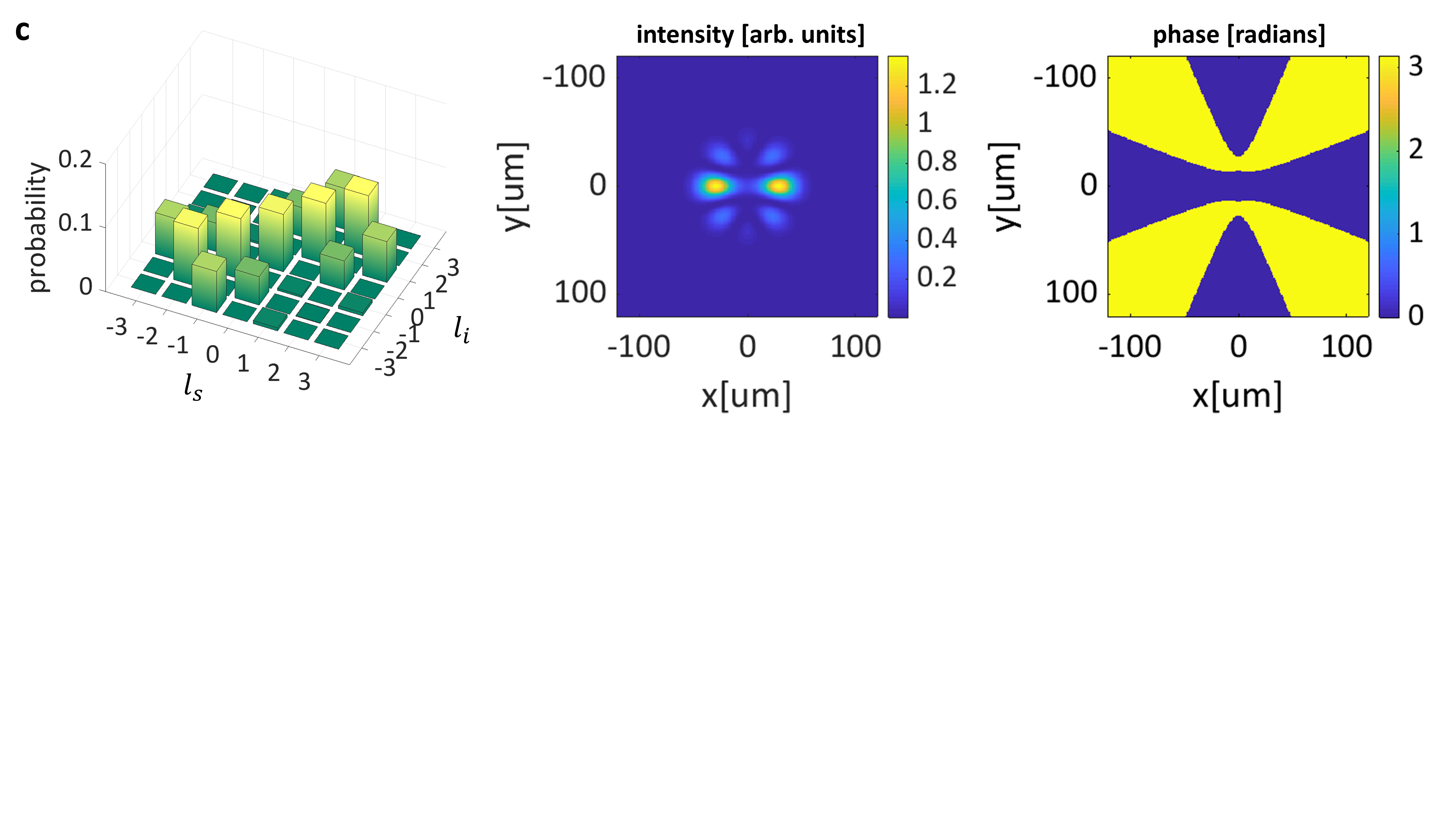} 
\end{tabular}
 \caption{ \small \textcolor{black}{Model validation against experimental results reported by Kovlakov et al. \cite{PhysRevA.98.060301}. \textbf{a} LG qutrit state: coincidence rate counts (left), pump intensity (middle) and phase (right). \textbf{b} Density matrix of the generated qutrit: real (left) and imaginary (right) parts of the density matrix. \textbf{c} LG ququint state: coincidence rate counts (left), pump intensity (middle) and phase (right).}}
 \label{fig:Kovlakov_PRA}
\end{figure*}

\begin{figure*}[]
\centering
\begin{tabular}{c}
    \includegraphics[width=0.95\linewidth]{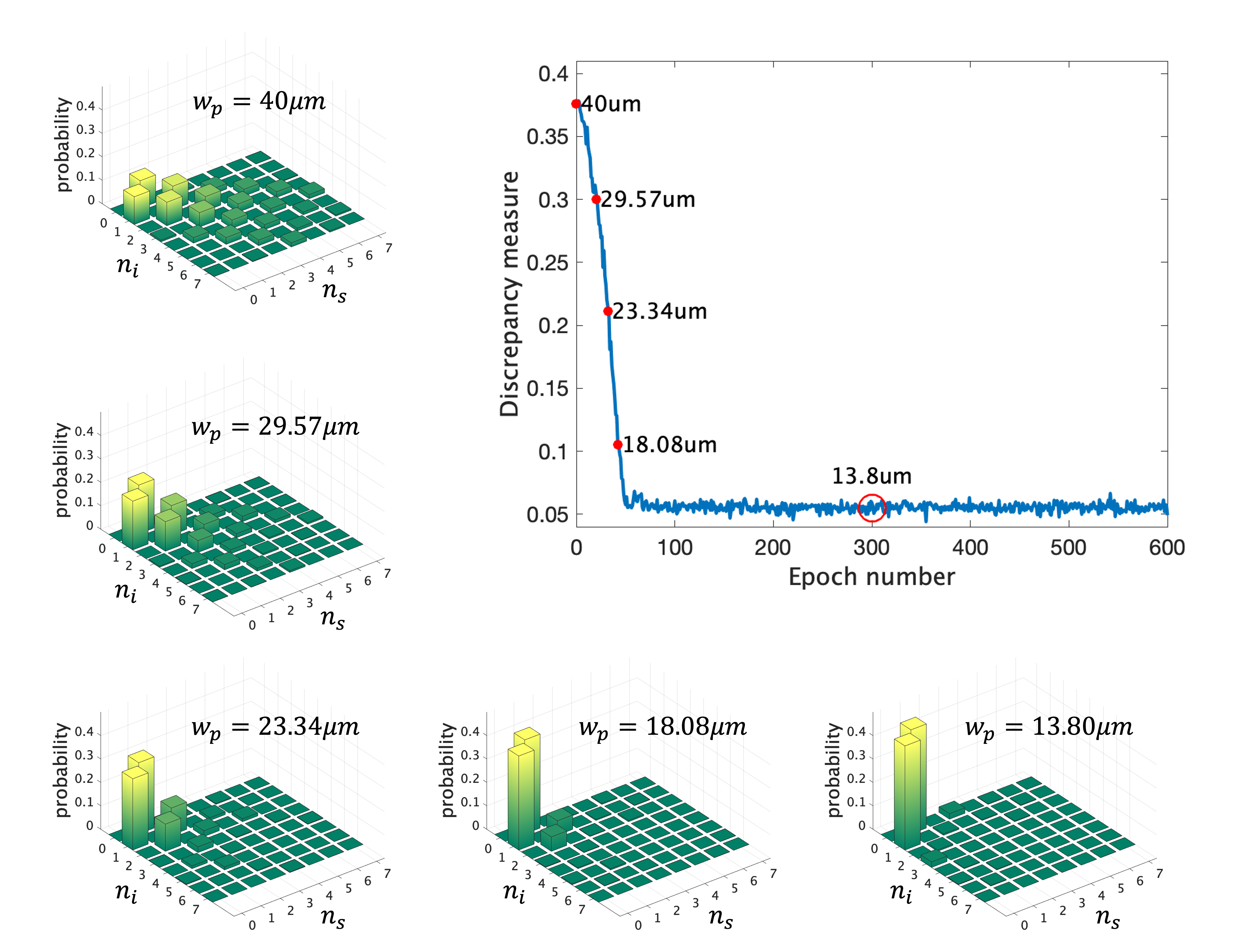}
\end{tabular}
 \caption{ \small \textcolor{black}{ Model validation against experimental results reported by Kovlakov et al. \cite{kovlakov2017spatial} for shaped correlations corresponding to the Bell state $(\ket{0,1}+\exp(i\phi)\ket{1,0})/\sqrt{2}$. The upper-right figure is the discrepancy measure (Eq. \ref{eq:optimizer}) between the generated coincidence rate counts and the desired one\cite{kovlakov2017spatial} vs training epoch number. The only learned physical parameter is the pump waist, and we let our algorithm find its optimal value for generating the desired quantum correlations. We sample the obtained pump waist along the discrepancy curve (red dots and insets) to see the evolution of the generated coincidence count rates under the optimized pump waist. At convergence, the algorithm obtains the correct pump waist value of $w_p=\sqrt{L/k_p}\approx13.8\mu m$ for $L=5mm$ for generating a pure HG Bell state.}}
 \label{fig:Kovlakov_PRL}
\end{figure*}

\textcolor{black}{
Before we delve into inverse design problems, we first validate our model against published experimental results of SPDC shaping \cite{kovlakov2017spatial, PhysRevA.98.060301}. This comes in addition to the multiple, already presented, validations of our model \cite{trajtenberg2020simulating}. Fig. \ref{fig:Kovlakov_PRA} presents the inference stage of our model for recovering the experimental results reported by Kovlakov et al. \cite{PhysRevA.98.060301}. We reproduce the coincidence rate counts for a qutrit state, Fig. \ref{fig:Kovlakov_PRA}a, and ququint state, Fig. \ref{fig:Kovlakov_PRA}c, in the LG basis, generated by a shaped pump field. To show the capability of our model to simulate the QST procedure, we recover the density matrix of the qutrit state, Fig. \ref{fig:Kovlakov_PRA}b, as reported by Kovlakov et al. \cite{PhysRevA.98.060301}. The resulting quantum states, coincidence rates and pump fields (used to recover the result in inference) are in good agreement with experiments (deviations may arise from detection, OAM projection, and coupling imperfections, as acknowledged by Kovlakov et al. \cite{PhysRevA.98.060301}). Next, we follow another result reported by Kovlakov et al. \cite{kovlakov2017spatial} and let our algorithm learn the optimal pump waist size for generating a pure HG spatial Bell state between structured SPDC photon pairs. Fig. \ref{fig:Kovlakov_PRL} shows the convergence of our learning algorithm towards the optimal pump waist, $w_p=\sqrt{L/k_p}$ \cite{kovlakov2017spatial}, for the case of $L=5 mm$. As the learning process progresses, the discrepancy measure, $\mathcal{D}(\cdot, \cdot)$ Eq. \ref{eq:optimizer}, reduces until the model reaches convergence. Accordingly, the size of the pump waist converges to the required value \cite{kovlakov2017spatial} and a clear Bell state, $(\ket{0,1}+\exp(i\phi)\ket{1,0})/\sqrt{2}$, is generated.}

\subsection{Shaping arbitrary quantum correlations}\label{subsec:results-G2}
First, we let our algorithm learn the physical parameters for desired quantum correlations -- that is, the two-photon coincidence rates -- between structured SPDC photon pairs. The learned parameters are the spatial modes of the crystal volume holograms and pump structure, according to Eq. \ref{eq:parameterized}. We use a type-2 SPDC process in a 1mm-long KTP NLPC, quasi-phase-matched to on-axis generation of photon pairs at 810nm from a 405nm pump wave. We assume that the pump beam is linearly polarized along the y direction and that the $\chi^{(2)}$ nonlinear coefficient can attain one of two binary values of $+d_{24}$ and $-d_{24}$. We project the generated photons on either the LG modes with the integer quantum numbers $l,p$, standing for the azimuthal and radial numbers, respectively; or the HG modes, with integer quantum numbers $n,m$, standing for the x and y-axis mode numbers, respectively. When considering the coincidence rate counts, we post-select either the radial index ($p=0$), in the case of LG basis, or the y-axis modal number ($m=0$), in the case of HG basis. The discrepancy measure in Eq. \ref{eq:optimizer} is taken as a weighted ensemble of the Kullback-Leibler divergence and the L1 norm.

\begin{figure*}[]
\centering
\begin{tabular}{c}
    \includegraphics[width=0.95\linewidth]{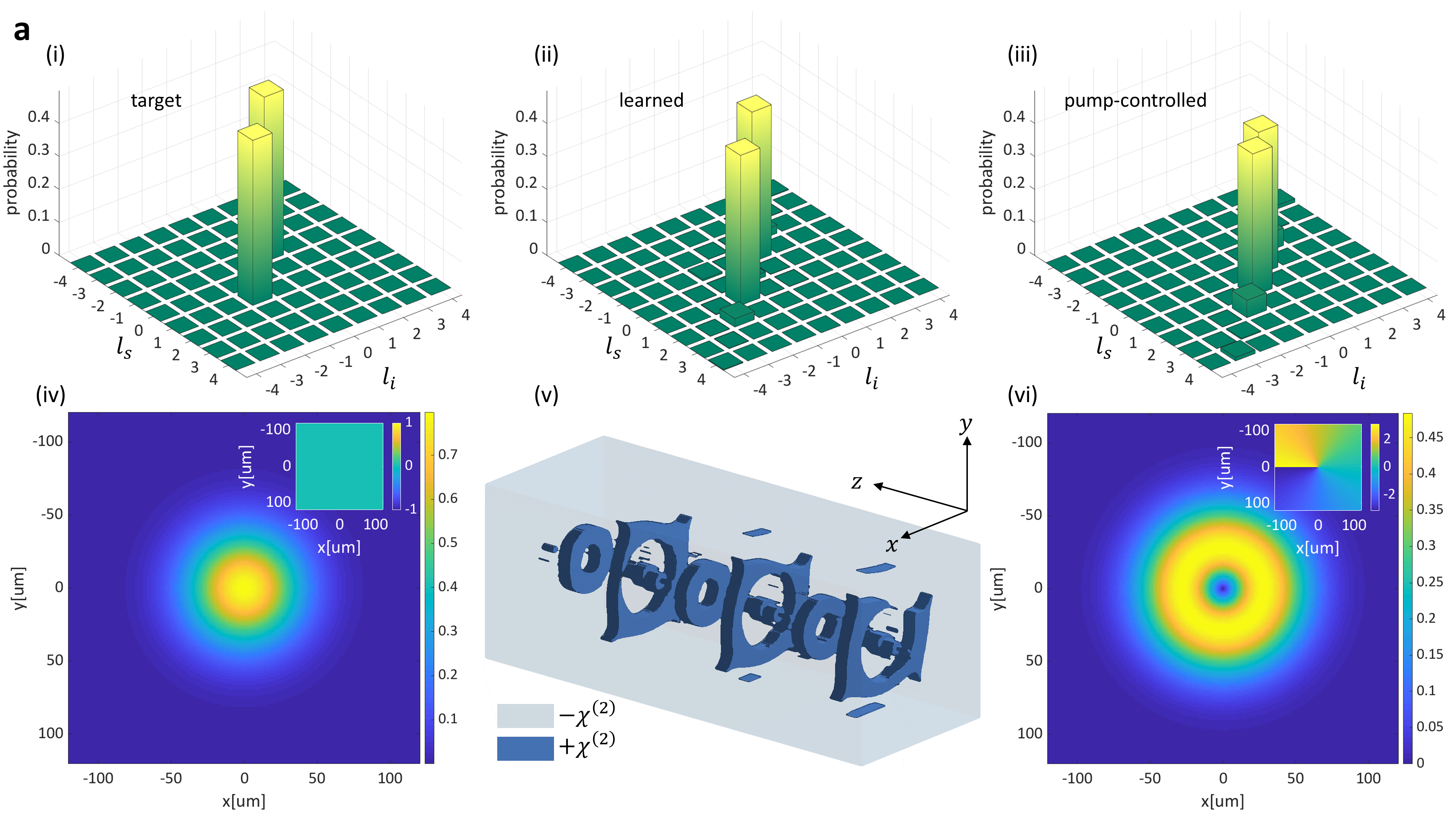}  \\
    
    \includegraphics[width=0.95\linewidth]{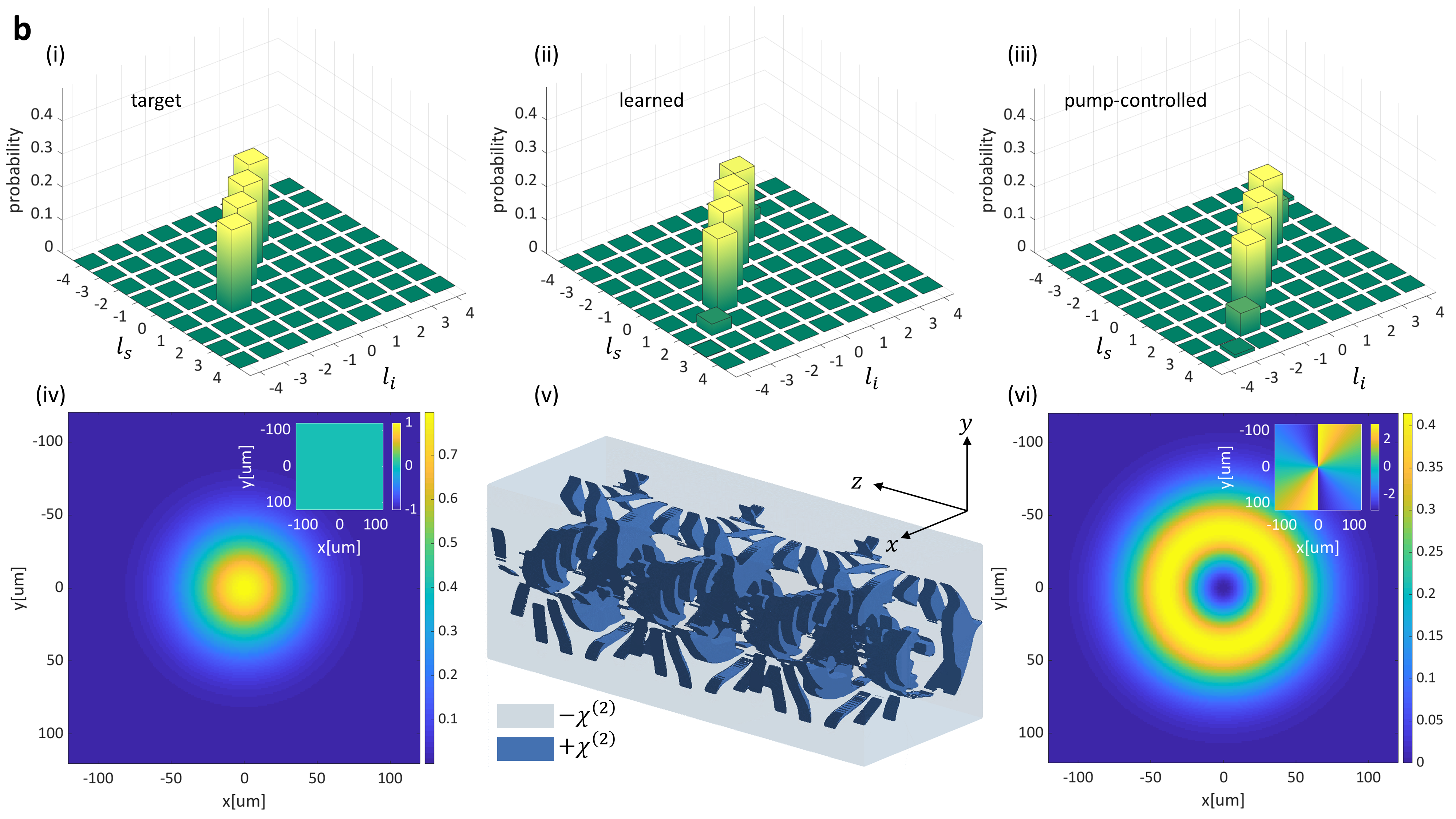}
\end{tabular}
 \caption{ \small Inverse design and all-optical coherent control over quantum correlations of SPDC photons: maximally-entangled two-photon states in the LG basis. \textbf{a}. Shaped correlations corresponding to the qubit state $\ket{\psi}=(\ket{1,-1}+\exp(i\phi)\ket{-1,1})/\sqrt{2}$. (i) shows the target coincidence probability. (ii) shows the learned coincidence probability, for an initial Gaussian pump (iv) and the learned 3D NLPC volume hologram (v). In (v), 3 successive unit cells are shown (the z-axis is scaled-up by a factor of 20). All-optical control over the coincidence probability is demonstrated using a $\mathrm{LG_{01}}$ pump mode (vi), with the same learned crystal -- giving quantum correlations that correspond to a new qubit state, $\ket{\psi}=(\ket{0,1}+\exp(i\phi)\ket{1,0})/\sqrt{2}$ (iii). \textbf{b}. Shaped correlations corresponding to the ququart state $\ket{\psi}=(\ket{-2,1}+\exp(i\phi_1)\ket{0,-1}+\exp(i\phi_2)\ket{-1,0}+\exp(i\phi_3)\ket{1,-2})/\sqrt{4}$. (i) to (v) as in \textbf{a}. All-optical control over the coincidence probability is demonstrated using a $\mathrm{LG_{02}}$ pump mode (vi), with the same learned crystal -- giving quantum correlations that correspond to a different ququart state, residing on the $l_i + l_s = +1$ diagonal, $\ket{\psi}=(\ket{2,-1}+\exp(i\phi_1)\ket{0,1}+\exp(i\phi_2)\ket{1,0}+\exp(i\phi_3)\ket{-1,2})/\sqrt{4}$ (iii).}
 \label{fig:lg1}
\end{figure*}

    
\paragraph{Laguerre-Gauss basis:}
Here, we show all-optically coherent control over quantum correlations of SPDC photons, in the LG basis (Fig. \ref{fig:lg1} depicts the results of this section). We use our algorithm to extract the optimal quantum volume holograms, embedded in 3D NLPCs, for generating the desired coincidence rate counts of maximally-entangled two-photon qubit $\ket{\psi}=(\ket{1,-1}+\exp(i\phi)\ket{-1,1})/\sqrt{2}$ and ququart $\ket{\psi}=(\ket{-2,1}+\exp(i\phi_1)\ket{0,-1}+\exp(i\phi_2)\ket{-1,0}+\exp(i\phi_3)\ket{1,-2})/\sqrt{4}$ states, that can later be actively-controlled via the pump beam (the indices of the signal and idler photons  are the azimuthal indices). We start by letting the algorithm learn the optimal 3D volume crystal hologram with a constant Gaussian pump beam, presented in Fig. \ref{fig:lg1}a(iv)-(v) and b(iv)-(v). The obtained volume holograms (Fig. \ref{fig:lg1}a-b(v)) display an intricate structures: concentric rings, Fig. \ref{fig:lg1}a(v), which mark the coupling to radial LG modes ($p>0$), and corkscrew structures, Fig. \ref{fig:lg1}b(v), indicating an intrinsic chirality of the hologram. We find that the coupling to radial modes is essential for quantum destructive and constructive interference in the post-selected subspace ($p=0$), while the crystal-handedness is responsible for inducing orbital angular momentum. The generated quantum correlations coincide remarkably well with the target, Fig. \ref{fig:lg1}a(i)-(ii) and b(i)-(ii)

The learned volume holograms demonstrate an even richer functionality -- they can span a larger variety of output correlations when the input pump mode is altered from Gaussian ($l=0$) to other LG modes, as depicted in Fig. \ref{fig:lg1}a-b(vi). As we alter the initial pump mode, the new correlations differ significantly from those obtained in the original design, while they still correspond to maximally-entangled states. Moreover, the new correlations keep the high signal to noise ratio (SNR) between the primary two-photon modes and the background of the coincidence signal, as can be seen in Figs. \ref{fig:lg1}a-b(iii). For example, by introducing an external pump orbital angular momentum, a qubit state originally on the $l_i + l_s = 0$ diagonal is shifted to a qubit on the $l_i + l_s = 1$ diagonal, Fig. \ref{fig:lg1}a, when $l_p = 1$. Similarly, a ququart on the $l_i + l_s = -1$ diagonal is shifted to the $l_i + l_s = 1$ diagonal when $l_p = 2$, Fig. \ref{fig:lg1}b. Interestingly, by using other learned holograms and superpositions of LG modes in the pump beam, we discover nontrivial pump-induced transformations, between a qutrit and a ququart, and a ququart and a qubit (see Supplementary Material, Section \ref{sup:lg}, Fig. \ref{fig:suppLG}).

\begin{figure*}[h]
\centering
\begin{tabular}{c}
    \includegraphics[width=0.95\linewidth]{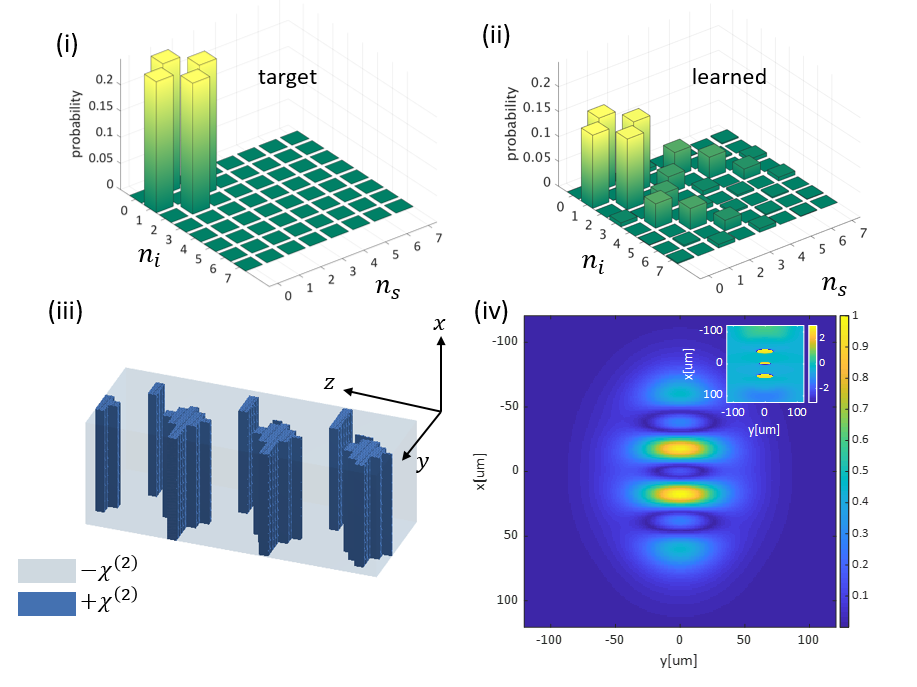}  

\end{tabular}
 \caption{ \small Inverse design of quantum correlations of SPDC photons: maximally-entangled two-photon states in the HG basis. Shaped correlations corresponding to the ququart state $\ket{\psi}=(\ket{0,1}+\exp(i\phi_1)\ket{1,0}+\exp(i\phi_2)\ket{1,2}+\exp(i\phi_3)\ket{2,1})/\sqrt{4}$. (i) and (ii) show, respectively, the target and learned coincidence rate counts. (iii) and (iv) show the simultaneously learned quantum volume hologram, embedded 2D NLPC, and complex pump beam profile, restricted to vary only in the x-direction. In (iii), 3 successive unit cells are shown (the z-axis is scaled-up by a factor of 20)} 
 \label{fig:hg}
\end{figure*}

\paragraph{Hermite-Gauss basis:}
In the previous example, in the Laguerre-Gauss basis, the learning step was performed by varying only the crystal parameter. Now, we show that by learning the quantum volume hologram and the pump beam profile, simultaneously, we can improve the quality of the generated second-order quantum correlations. In this section, we explore the photon correlations in the HG basis and our target is a two-photon ququart state $\ket{\psi}=(\ket{0,1}+\exp(i\phi_1)\ket{1,0}+\exp(i\phi_2)\ket{1,2}+\exp(i\phi_3)\ket{2,1})/\sqrt{4}$ (the indices of the signal and idler photons are the Hermite-Gaussian modes indices in the Y direction). We consider designs that use more mature NLPC technologies, such as electric field poling \cite{kazansky1997electric}, which are restricted to 2D nonlinear holograms. We use our algorithm to simultaneously extract the optimal quantum volume hologram that varies only in the y-direction (embedded in 2D NLPC) and the pump beam profile that is restricted to vary only in the x-direction, for generating the desired coincidence rate counts of a maximally-entangled two-photon ququart. In Fig. \ref{fig:hg}(ii) we see the generated coincidence rate counts that result from the computed interaction parameters. While the probabilities of the generated ququart state are lower than the desired target, they are equal and significantly larger than other unwanted probabilities. This result is certainly exciting when taking into account the restrictions we considered under 2D-variation. The obtained volume hologram (Fig. \ref{fig:hg}(iii)) and the pump profile (Fig. \ref{fig:hg}(iv)) display a Cartesian structure.

To better show the importance of combining both the quantum volume hologram and the pump beam profile to obtain the desired maximally-entangled state, we compared the quality of the generated second-order quantum correlations of a ququart state under the following three scenarios: using our algorithm to solve the inverse design problem and 1) extracting the optimal quantum volume hologram, embedded in 3D NLPC, with a constant Gaussian pump; 2) extracting the complex pump beam profile, with a constant periodically-poled crystal; 3) extracting both the optimal quantum volume hologram, embedded 3D NLPC, and the optimal complex pump beam profile. The simultaneous learning of the pump and crystal clearly outperform the individual learning of either. This is attributed to higher modes created by the multiplication of modes composing the pump and crystal structure in the nonlinear coupling coefficient, $\kappa_j$ (in Eq. \ref{eq:waveeq}). Also, there seemed to be no preference in the generated results while optimizing separately either the NLPC or the pump, which shows the similar role of each of them in the nonlinear coupling coefficient. For visual results, see Supplementary Material, section \ref{sup:hg}, Fig. \ref{fig:suppHG}.

\subsection{Shaping arbitrary quantum states}
In order to resolve a specific two-photon quantum state generated by the tailored SPDC process, a coincidence measurement will not suffice. For this purpose, we emulate QST and integrate it into our learning stage for evaluating the corresponding density matrix, as detailed in Section \ref{subsec:observables}. The density matrix is used as an observable while the Trace Distance is taken as the discrepancy metric $\mathcal{D}(\cdot, \cdot)$ (Eq. \ref{eq:optimizer}). As a proof-of-concept, we consider two-photon qudit states with dimension $d=3$ in the LG basis. That is, we focus on the subspace spanned by $\lbrace\ket{-1}, \ket{0}, \ket{1} \rbrace\otimes\lbrace\ket{-1}, \ket{0}, \ket{1}\rbrace$, giving a 9-by-9 dimensional density matrix.

\begin{figure*}[]
\centering
\begin{tabular}{c}
    \includegraphics[width=0.95\linewidth]{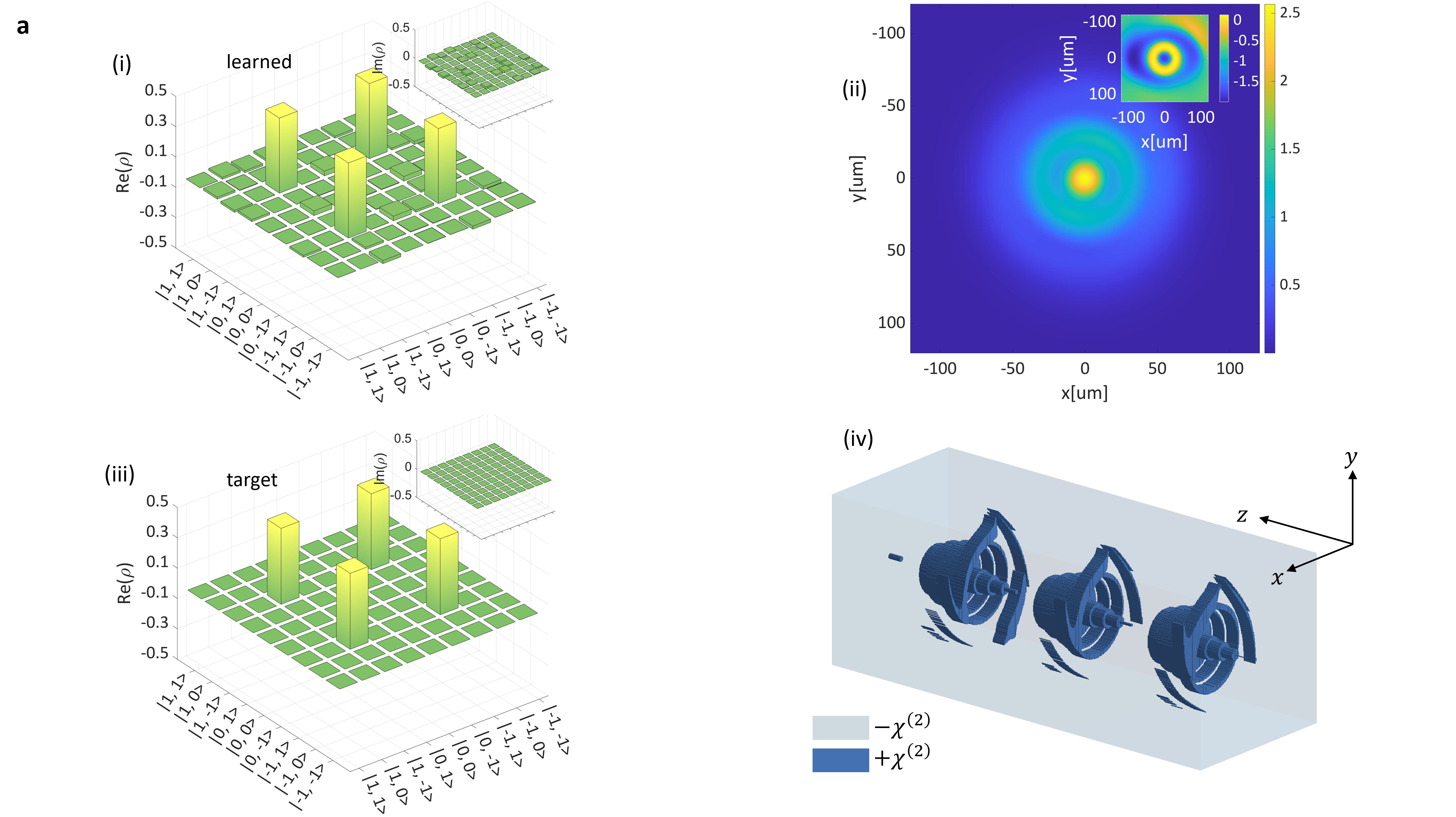}  \\
    
    \includegraphics[width=0.95\linewidth]{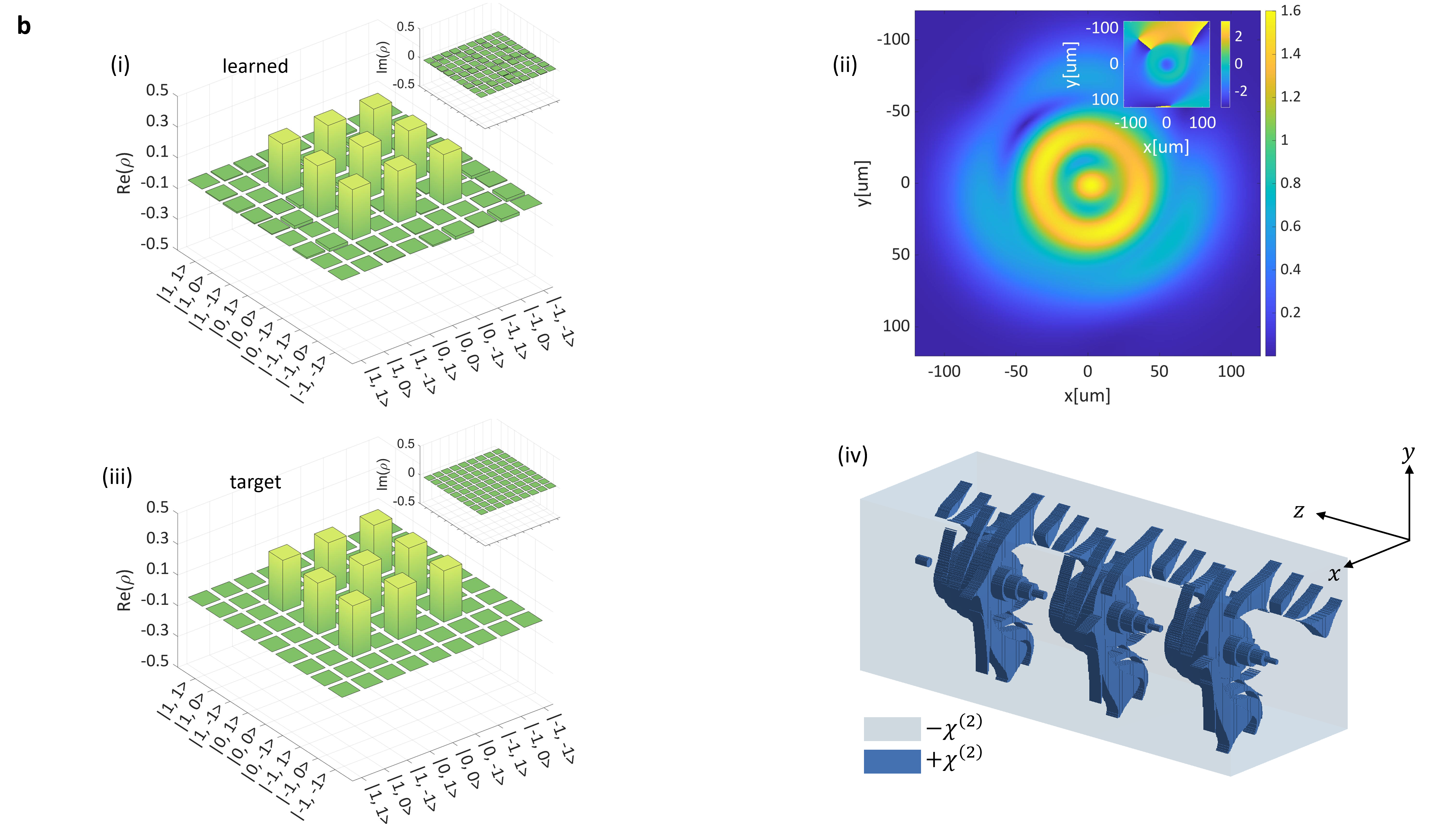}
\end{tabular}
\caption{ \small Inverse design of quantum state density matrices of SPDC photons: maximally-entangled two-photon states in the LG basis.  \textbf{a}. The qubit state $\ket{\psi}=(\ket{1,-1}+\ket{-1,1})/\sqrt{2}$. (i) and (iii) show, respectively, the learned and target states (the real part of the density matrix is shown in large, and the imaginary in small). (ii) and (iv) show the simultaneously learned complex pump beam profile and quantum volume hologram embedded 3D NLPC. In (iv), 3 successive unit cells are shown (the z-axis is scaled-up by a factor of 20). \textbf{b}. The qutrit state $\ket{\psi}=(\ket{1,-1}+\ket{0,0}+\ket{-1,1})/\sqrt{3}$. (i-iv) as in \textbf{a}.}
 \label{fig:rho1}
\end{figure*}

Similar to the previous subsection, we use our algorithm to simultaneously extract the optimal quantum volume holograms, embedded in 3D NLPCs, and the pump beam profiles, for generating the desired quantum states. Fig. \ref{fig:rho1}a depicts the results for the maximally-entangled state $\ket{\psi}=(\ket{1,-1} + \ket{-1,1})/\sqrt{2}$ (corresponding to the coincidence rate shown in Fig. \ref{fig:lg1}a(i)), while Fig. \ref{fig:rho1}b depicts the results for the maximally-entangled state $\ket{\psi}=(\ket{1,-1} + \ket{0,0} +\ket{-1,1})/\sqrt{3}$ (corresponding to the coincidence rate shown in the Supplementary Material, Fig. \ref{fig:suppLG}a(i)). The generated density matrices fit the target states well, as evident in Figs. \ref{fig:rho1}a(i),(iii) and b(i),(iii). Our learned pump profiles and crystal holograms demonstrate concentric shapes, Fig. \ref{fig:rho1}a(ii),(iv) and b(ii),(iv). These maintain a total orbital angular momentum of $l_i + l_s = 0$, as expected, while making higher-order radial LG modes possible. These higher order modes are responsible, for example, for removing the two-photon Gaussian mode $\ket{00}$ in the first learned state, Figs. \ref{fig:rho1}a(i) and \ref{fig:lg1}a(ii), through destructive interference, which is otherwise impossible when only using Gaussian pump beams.

Importantly, the generated quantum two-photon states are sensitive to the relative phase between the modes constructing the pump profile and the learned nonlinear volume holograms. This feature is essential for asserting that the active all-optical control over the coincidence rates, discussed in the previous section, allows also for quantum coherent control over the generated photon qudits. To demonstrate this, we again learn a 3D volume hologram with a fixed pump profile, but this time consisting of a given superposition of LG modes. By changing the relative phase between the LG modes, we expect that the off-diagonal terms in the density matrix change accordingly. 

\begin{figure*}[h]
\centering
\begin{tabular}{c}

    \includegraphics[width=0.95\linewidth]{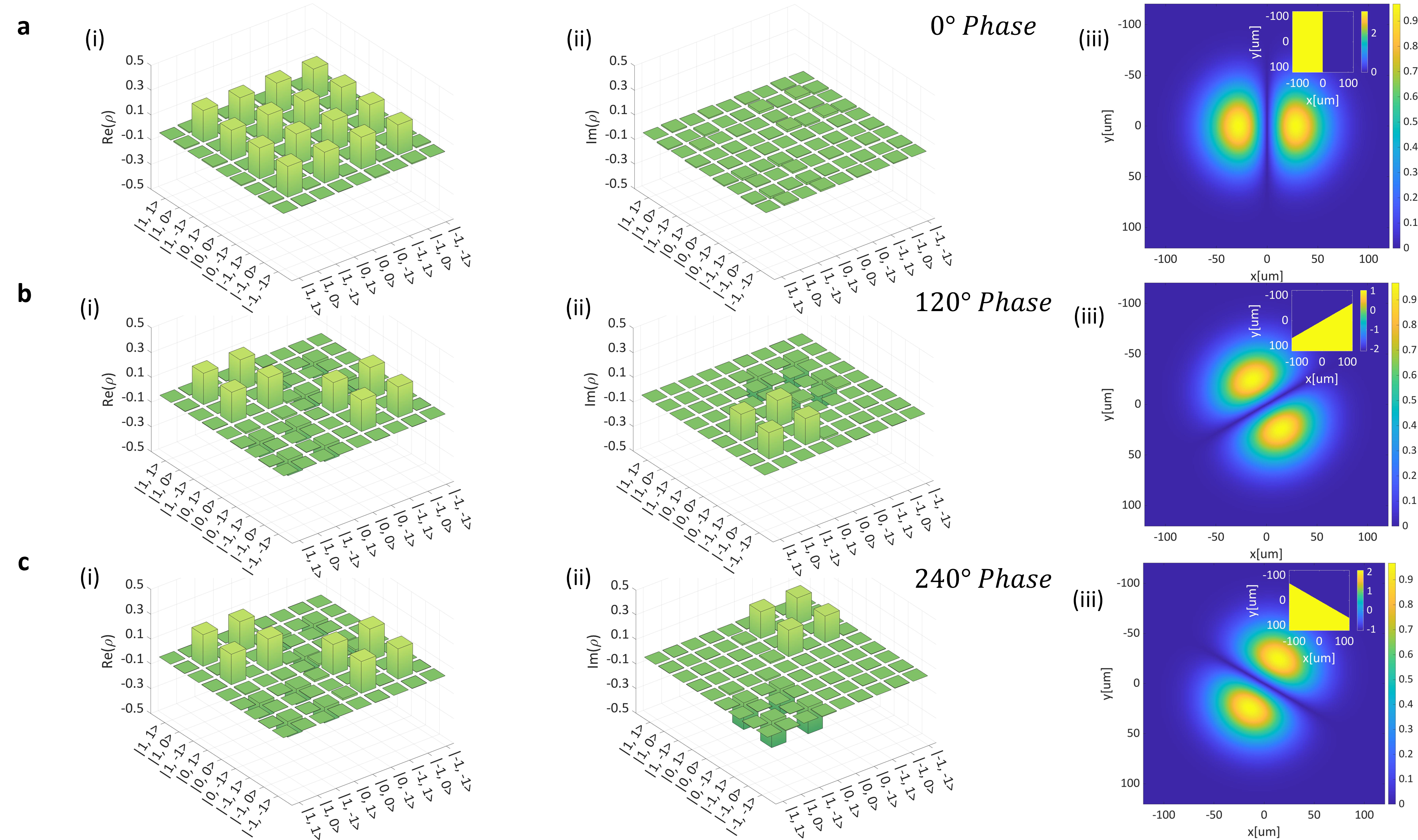}
\end{tabular}
 \caption{ \small Inverse design and all-optical coherent control over quantum state of SPDC photons: maximally-entangled two-photon ququart state in the LG basis. We use our algorithm to extract the 3D NLPC hologram that generates the desired ququart state $\ket{\psi}=(\ket{-1,0}+\ket{0,-1}+\ket{1,0}+\ket{0,1})/\sqrt{4}$, using the initial constant pump profile $\mathrm{HG_{10}} = \mathrm{LG_{01}}+\mathrm{LG_{0-1}}$ 
 a(iii). The real part of generated density matrix is shown in a(i) and the imaginary part in a(ii). Next, the pump beam illuminating the learned hologram is rotated to actively control the generated quantum state. b(i) and (ii) show the real and imaginary parts, respectively, of generated density matrix for the rotated incident beam $\mathrm{LG_{01}}+e^{i120^{\circ}}\mathrm{LG_{0-1}}$ b(iii). c(i)-(ii) show the real and imaginary parts, respectively, of generated density matrix for the rotated incident beam $\mathrm{LG_{01}}+e^{i240^{\circ}}\mathrm{LG_{0-1}}$ c(iii). }
 \label{fig:rho2}
\end{figure*}

Fig. \ref{fig:rho2} depicts the results for the generated maximally-entangled two-photon ququart state $\ket{\psi}=(\ket{-1,0}+\ket{0,-1}+\ket{1,0}+\ket{0,1})/\sqrt{4}$. Initially, we use our algorithm to extract the optimal quantum volume hologram, embedded in 3D NLPCs, with a fixed pump beam of the form: $\mathrm{LG_{01}}+e^{i\alpha}\mathrm{LG_{0-1}}$ for $\alpha=0^{\circ}$ (i.e., a $\mathrm{HG}_{10}$ mode, as presented in Fig. \ref{fig:rho2}a(iii)). The real part of the generated density matrix is shown in Fig. \ref{fig:rho2}a(i) and the imaginary part in Fig. \ref{fig:rho2}a(ii). The generated density matrix fits the desired one. We then used the extracted crystal volume hologram with different superpositions of LG modes of the pump. Figs. \ref{fig:rho2}b(i)-(ii) and c(i)-(ii) show the quantum states achieved through inference with the same learned crystal hologram, but with the pump mode superposition phase angle $\alpha$ changed to $\alpha=120^{\circ}$, Fig. \ref{fig:rho2}b(iii), and $240^{\circ}$, Fig. \ref{fig:rho2}c(iii). This corresponds experimentally to a rotation of the $\mathrm{HG}_{10}$ mode. Note, the significant change in the imaginary off-diagonal density matrix elements, in Figs. \ref{fig:rho2}b(ii) and c(ii). This indicates the coherent control over the quantum state via the rotation of the pump beam -- a diverse functionality available by use of a single volume hologram pumped with different optical modes. 

\section{Conclusion} \label{sec:conclusion}
We have introduced an algorithm for solving the inverse design problem of generating structured and entangled photon pairs in quantum optics, using tailored nonlinear interactions in the SPDC process. The \emph{SPDCinv} algorithm extracts the optimal physical parameters which yield a desired quantum state or correlations between structured photon-pairs, that can then be used in future experiments. To ensure convergence to  realizable results and to improve the predictive accuracy, our algorithm obeyed physical constraints through the integration of the time-unfolded propagation dynamics governing the interaction of the SPDC Hamiltonian. We have shown how we can apply our algorithm to obtain the optimal nonlinear $\chi^{(2)}$ volume holograms (2D/3D) as well as different pump structures for generating the desired maximally-entangled states. \textcolor{black}{The optimal crystal holograms extracted by our model seem to exhibit robustness against imperfections. To mimic crystal fabrication imperfections we deliberately add errors to the crystal structure to impair the generated coincidence rate counts of the maximally-entangled two-photon qubit.  Then, we show how with a slight variation in a different parameter of the system (pump waist), we can divert the system back, to nearly recover the original system results (see supplementary section \ref{sup:imperfections}).} The high dimensionality of these generated states increases the bandwidth of quantum information, and can improve the security of quantum key distribution protocols \cite{fuchs1997optimal,durt2004security}. We further demonstrate all-optical coherent control over the generated quantum states by actively changing the profile of the pump beam, making our results appealing for a variety of quantum information applications that require fast switching rates.

This work can readily be extended to the spectral-temporal domain, by allowing non-periodic volume holograms along the propagation axis -- making it possible to shape the joint spectral amplitude \cite{zielnicki2018joint} of the photon pairs. Furthermore, one can easily adopt our approach for other optical systems, such as: nonlinear waveguides and resonators \cite{QiLi+2020+1287+1320}, $\chi^{(3)}$ effects (e.g. spontaneous four wave mixing \cite{sharping2006generation}), spatial solitons \cite{stegeman1999optical, chen2012optical}, fiber optics communication systems \cite{10.1007/3-540-46629-0_9, hager2020physics}, and even higher-order coincidence probabilities \cite{PhysRevA.78.033831}. Moreover, the algorithm can be upgraded to include passive optical elements such as beam-splitters, holograms, and mode sorters \cite{krenn2016automated}, thereby providing greater flexibility for generating and manipulating quantum optical states. \textcolor{black}{Our model can incorporate decoherence mechanisms arising from non-perturbative high-order photon pair generation in the high gain regime \cite{brambilla2004simultaneous, trajtenberg2020simulating}. Other decoherence effects due to losses such as absorption and scattering can be incorporated into the model in the future.} Finally, our current scheme can be adapted to other quantum systems sharing a similar Hamiltonian structure, such as superfluids and superconductors \cite{coleman2015introduction}. In light of all this, we believe that this work, along with its complementary code, can contribute to further exciting advancements and discoveries in other quantum and classical systems.




\bibliography{references} \label{sec:references}

\appendix
\counterwithin{figure}{section}
\section{Supplementary Material} \label{sec:supplementary}


In this section we will show a nontrivial all-optical coherent control, between a qutrit and a ququart, and a ququart and a qubit. We will also present the importance of combining both the quantum volume hologram and the pump beam structure to obtain more accurate results, by comparing different scenarios.

\subsection{Shaping and all-optical control over quantum correlations in the LG basis} \label{sup:lg}

\begin{figure*}[]
\centering
\begin{tabular}{c}
    
    \includegraphics[width=0.95\linewidth]{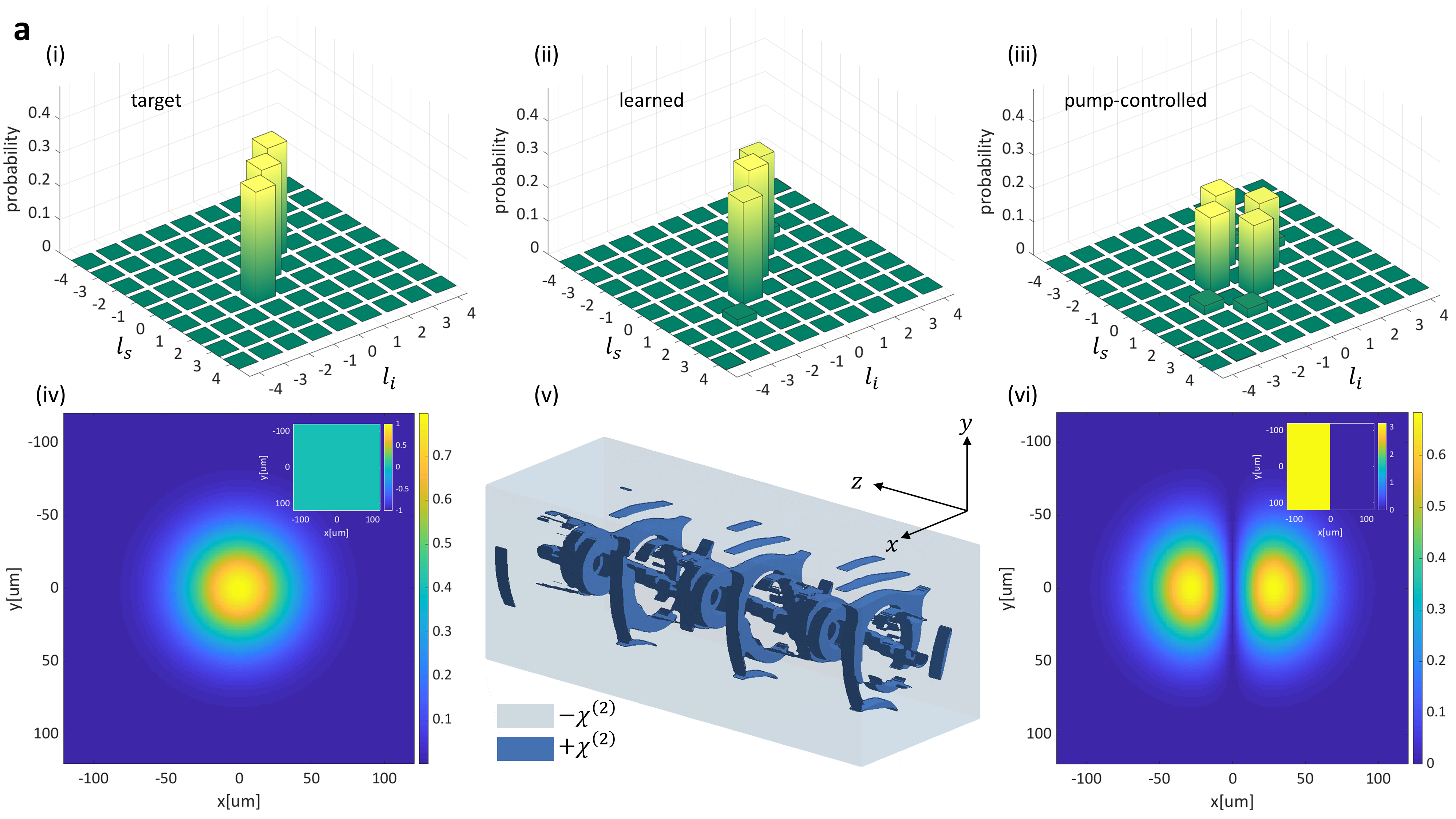} \\
    
    \includegraphics[width=0.95\linewidth]{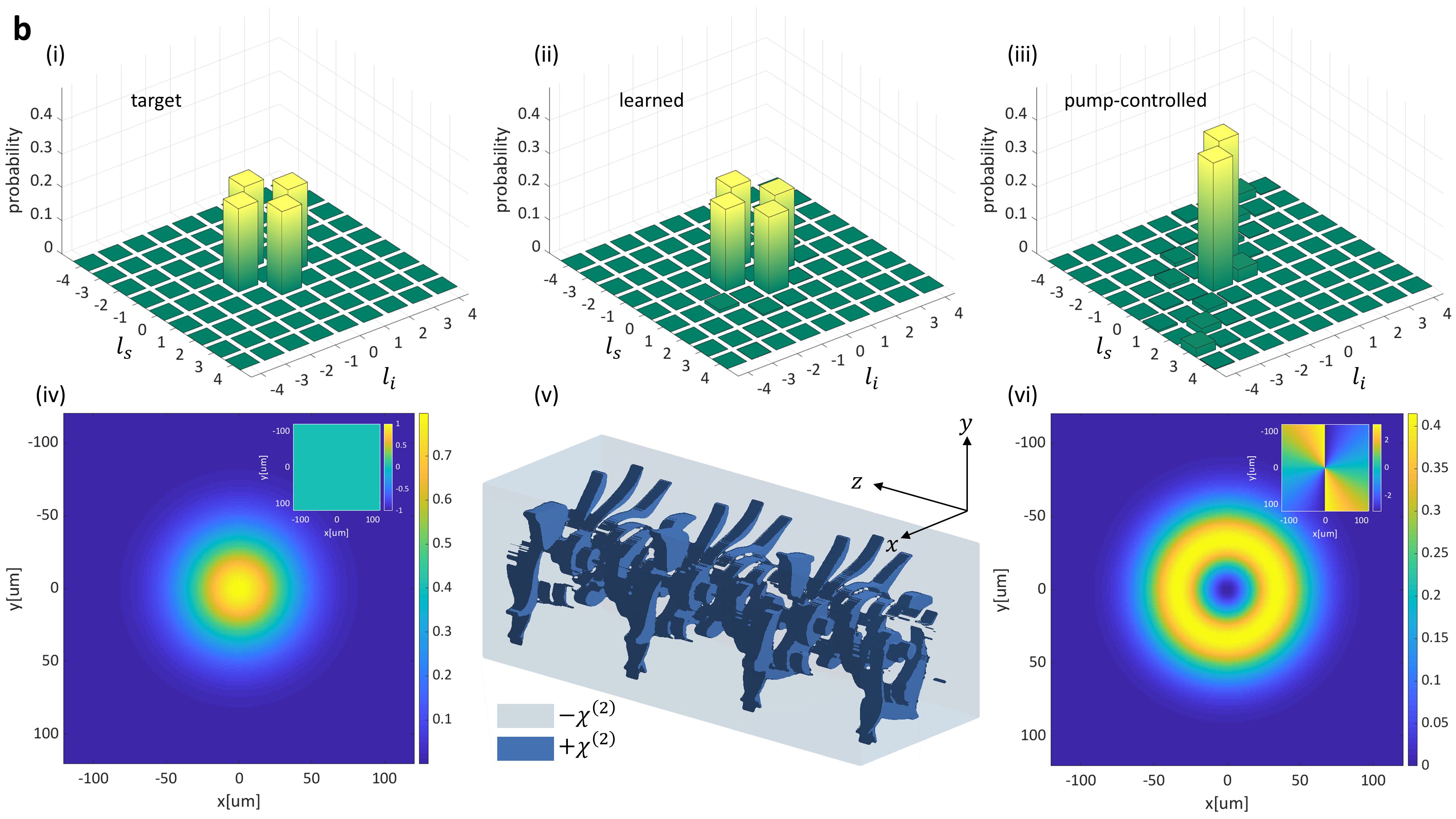}
\end{tabular}
\caption{ \small Inverse design and all-optical coherent control over quantum correlations of SPDC photons: maximally-entangled two-photon states in the LG basis. \textbf{a}. Changing entanglement by utilizing Hermite-Gauss pump beam. We begin by generating shaped correlations corresponding to the qutrit state $\ket{\psi}=(\ket{1,-1}+\ket{0,0}+\ket{-1,1})/\sqrt{3}$. (i) shows the target coincidence probability. (ii) shows the learned coincidence probability, for an initial Gaussian pump (iv) and the learned 3D NLPC volume hologram (v). In (v), 3 successive unit cells are shown (the z-axis is scaled-up by a factor of 20). All-optical control over the coincidence probability is demonstrated using a $\mathrm{HG_{10}}$ pump mode (vi), with the same learned crystal -- giving quantum correlations that correspond to a ququart state $\ket{\psi}=(\ket{1,0}+\ket{0,1}+\ket{-1,0}+\ket{0,-1})/\sqrt{4}$ (iii). \textbf{b}. Destructive interference using higher-order Laguerre-Gauss radial modes. We begin by generating shaped correlations corresponding to the ququart state $\ket{\psi}=(\ket{1,0}+\ket{0,1}+\ket{-1,0}+\ket{0,-1})/\sqrt{4}$. (i) to (v) as in \textbf{a}. All-optical control over the coincidence probability is demonstrated using a $\mathrm{LG_{02}}$ pump mode (vi), with the same learned crystal -- giving quantum correlations that correspond to a qubit state $\ket{\psi}=(\ket{-1,0}+\ket{0,-1})/\sqrt{2}$ (iii).}
 \label{fig:suppLG}
\end{figure*}

In the main text we showed that changing the LG pump modal number induces a shift of the quantum state (Fig. \ref{fig:lg1}a-b in the main text).  In addition, we saw that the use of an HG pump with different orientations provides all-optical coherent control over the two-photon state (Fig. \ref{fig:rho2} in the main text). In this section, we show further examples for all-optically coherent control over quantum correlations of SPDC photons, in the LG basis  (Fig. \ref{fig:suppLG} depicts the results of this section). We start by letting our algorithm extract the optimal quantum volume holograms, embedded in 3D NLPCs, for generating the desired coincidence rate counts of the maximally-entangled two-photon qutrit state $\ket{\psi}=(\ket{1,-1}+\ket{0,0}+\ket{-1,1})/\sqrt{3}$, with a constant Gaussian pump beam, presented in Fig. \ref{fig:suppLG}a(iv)-(v). The generated quantum correlations coincide well with the target, Fig. \ref{fig:suppLG}a(i)-(ii). The initial Gaussian pump is then replaced with an $\mathrm{HG_{10}}$ pump, Fig. \ref{fig:suppLG}a.(vi), and projected on the learned 3D hologram, \ref{fig:suppLG}a(v). The resulting coincidence rate, interestingly, is that of the maximally entangled ququart state $\ket{\psi}=(\ket{1,0}+\ket{0,1}+\ket{-1,0}+\ket{0,-1})/\sqrt{4}$, Fig. \ref{fig:suppLG}a(iii). This represents a change in the nature of the entangled photon states (e.g., transforming it from a biphoton qutrit to a biphoton ququart), demonstrating an additional functionality of the 3D NLPC holograms driven by different optical pumps. It is evident that the new correlations keep the high SNR.

In the main text, the constructive and destructive interference induced by the high-order LG radial modes was mentioned. We discussed it in section \ref{subsec:results-G2} when exploring quantum correlations in the LG basis. The following example reinforces that statement. Again, we start by letting our algorithm extract the optimal quantum volume holograms, embedded in 3D NLPCs, for generating the desired coincidence rate counts of maximally-entangled two-photon ququart state $\ket{\psi}=(\ket{1,0}+\ket{0,1}+\ket{-1,0}+\ket{0,-1})/\sqrt{4}$, with a constant Gaussian pump beam, presented in Fig. \ref{fig:suppLG}b(iv)-(v). The generated quantum correlations coincide well with the target, Fig. \ref{fig:suppLG}b(i)-(ii). The initial Gaussian pump is then replaced with an $\mathrm{LG_{02}}$ pump, Fig. \ref{fig:suppLG}b.(vi), and projected on the learned 3D hologram, Fig. \ref{fig:suppLG}b(v). We now see that the quantum correlations on the $l_i + l_s = 1$ diagonal disappear, thanks to the destructive interference created by the new incident pump beam. The resulting coincidence rate is that of a maximally entangled qubit state $\ket{\psi}=(\ket{-1,0}+\ket{0,-1})/\sqrt{2}$, Fig. \ref{fig:suppLG}b.(iii) and (vi). In this manner, we enable a non-trivial switching between ququart and qubit states.

\newpage
\subsection{Comparison of inverse design performance using pump-only, hologram-only, and pump-and-hologram learning} \label{sup:hg}

\begin{figure*}[]
\centering
\begin{tabular}{c}
    
    \includegraphics[width=0.95\linewidth]{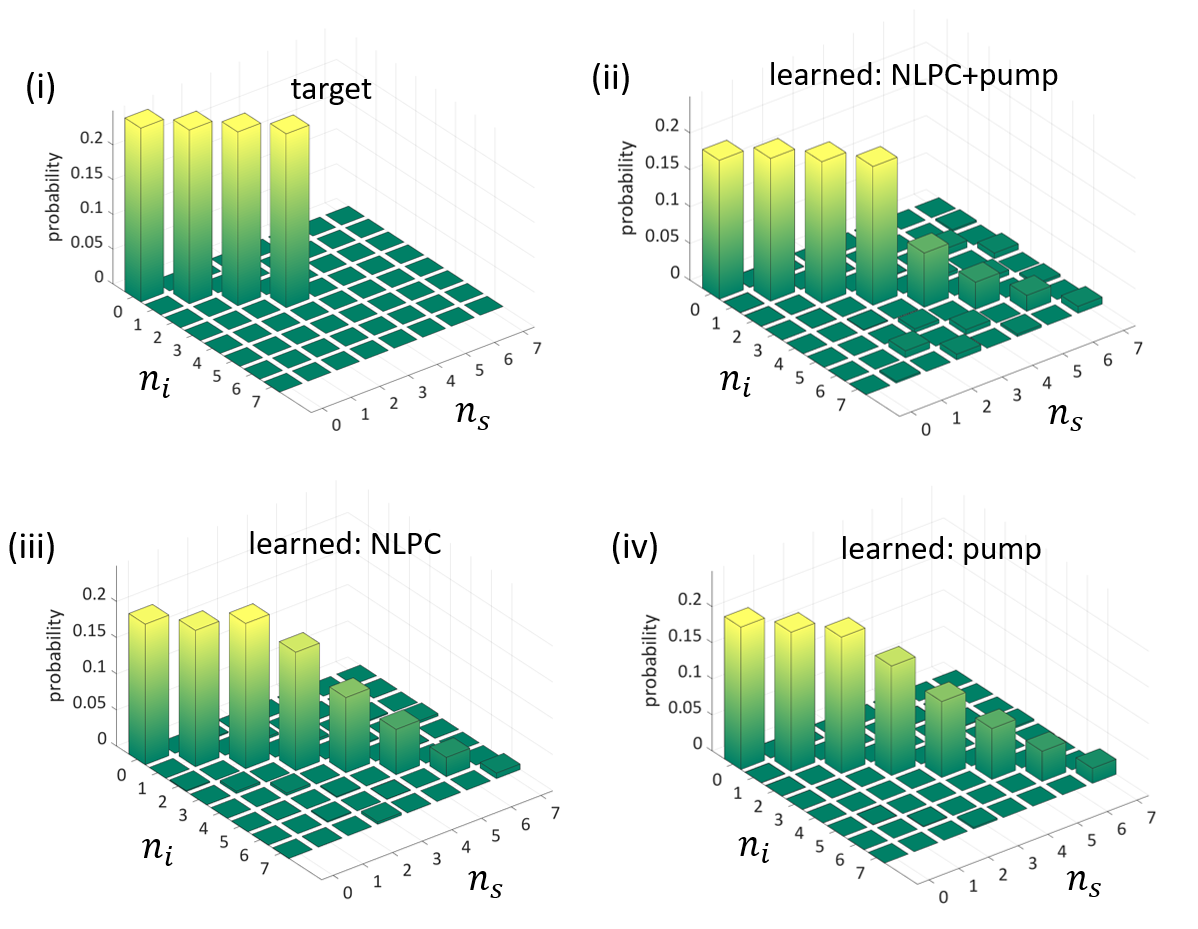}
    
\end{tabular}
 \caption{ \small Comparing the inverse design performance under different learning methods for generating second-order quantum correlations, in the HG basis, corresponding to the ququart state $\ket{\psi}=(\ket{0,0}+\exp(i\phi_1)\ket{1,1}+\exp(i\phi_2)\ket{2,2}+\exp(i\phi_3)\ket{3,3})/\sqrt{4}$. The desired second-order quantum correlations are given in (i). In (ii), we see the generated second-order quantum correlations while letting our algorithm solve the inverse design problem and extract both the optimal quantum volume hologram, embedded 3D NLPC and the optimal complex pump beam structure. In (iii), we see the generated second-order quantum correlations while letting our algorithm solve the inverse design problem and extract the optimal quantum volume hologram, embedded in the 3D NLPC, with a constant Gaussian pump. In (iv),  we see the generated second-order quantum correlations while letting our algorithm solve the inverse design problem and extract the optimal complex pump beam structure, with a constant periodically-poled crystal.}
 \label{fig:suppHG}
\end{figure*}

Despite the functionality provided by the learned volume holograms, simultaneous learning of both the optimal quantum volume hologram ,embedded in NLPC, and the pump beam structure will usually outperform individual learning of either. As can be seen in Fig. \ref{fig:suppHG}, we compared the quality of the generated second-order quantum correlations, in the HG basis, of a ququart state under the following three scenarios: 1) We use our algorithm to solve the inverse design problem and extract both the optimal quantum volume hologram and the optimal complex pump beam structure, Fig. \ref{fig:suppHG}(ii). 2) We use our algorithm to solve the inverse design problem and extract the optimal quantum volume hologram, with a constant Gaussian pump, Fig. \ref{fig:suppHG}(iii). 3) We use our algorithm to solve the inverse design problem and extract the optimal complex pump beam structure, with a constant periodically-poled crystal, Fig. \ref{fig:suppHG}(iv). The desired second-order quantum correlations are given in Fig. \ref{fig:suppHG}(i).

The state achieved through pump and crystal learning, Fig. \ref{fig:suppHG}(ii), not only achieves more equal amplitudes for the modes of the ququart state, but also leaks less energy to unwanted modes. This can be explained by the higher order modes achieved through the multiplication of the pump and crystal modes in the nonlinear coupling coefficient, $\kappa_j$ (in Eq. \ref{eq:waveeq}). Also, as the pump and crystal are allowed to learn the same set of modes, their learned results seems very similar. It appears to be no preference in the generated results while optimizing them separately, which shows the similar role of each of them in the nonlinear coupling coefficient, $\kappa_j$.

\newpage
\subsection{\textcolor{black}{Effects of crystal imperfections}} \label{sup:imperfections}
\begin{figure*}[!ht]
\centering
\begin{tabular}{c}
    
    \includegraphics[width=0.95\linewidth]{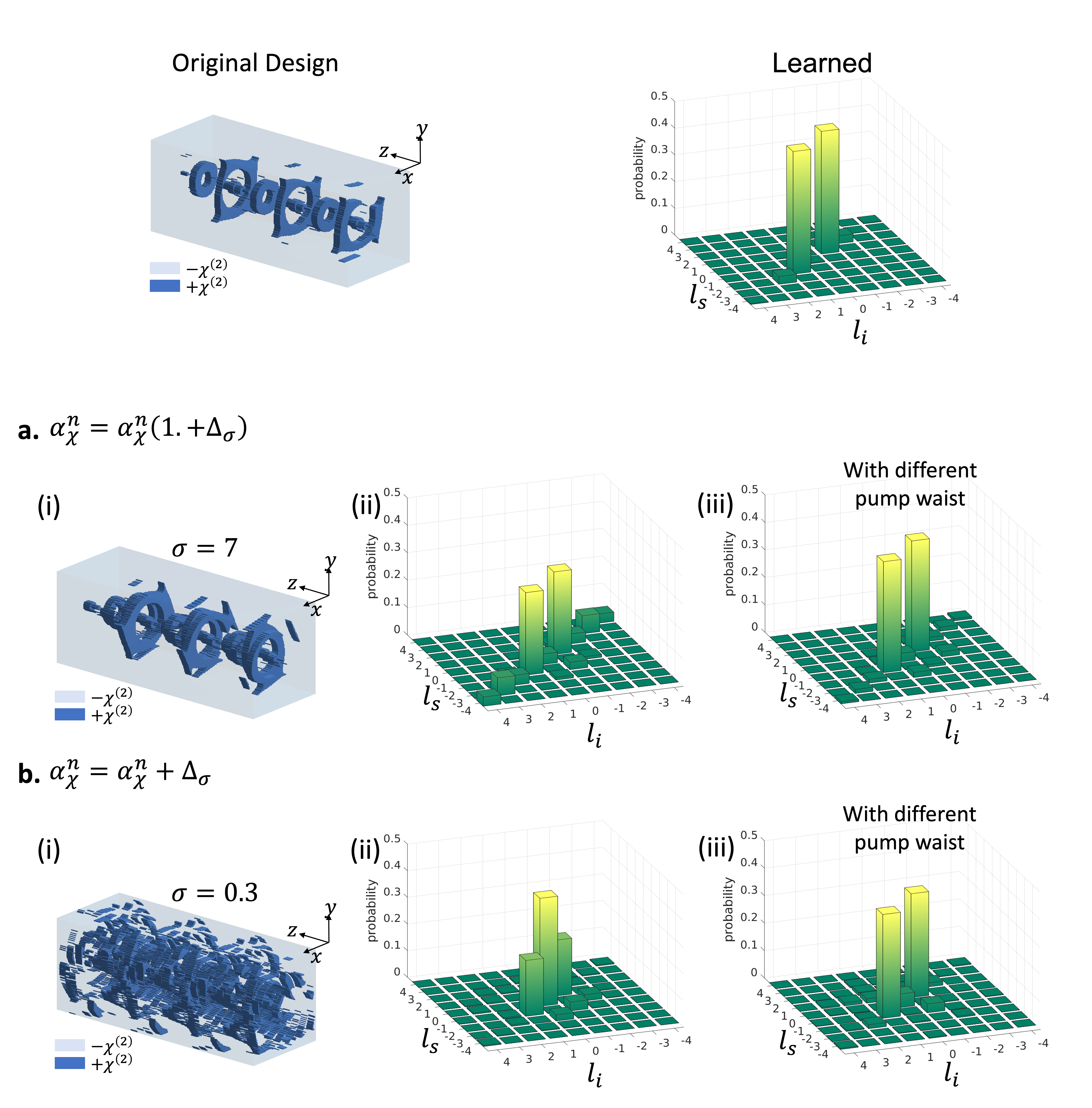}
    
\end{tabular}
 \caption{ \small \textcolor{black}{Tolerance of generated coincidence rate counts of maximally-entangled two-photon qubit $\ket{\psi}=(\ket{1,-1}+\exp(i\phi)\ket{-1,1})/\sqrt{2}$, under imperfect 3D volume crystal hologram structure. The original design and corresponding learned result were taken from the main text for convenient visual comparison with the current findings (Figs. \ref{fig:lg1}.a(v) and 5.a(ii)). a-b(i) show two noisy versions of the optimal quantum volume hologram (Fig. \ref{fig:lg1}.a(v)) for generating the coincidence rate counts of the desired qubit state. In a-b(i), 3 successive unit cells are shown (the z-axis is scaled-up by a factor of 20). a-b(ii) show the impaired coincidence rate counts, compare to the original setup (Fig. \ref{fig:lg1}.a(ii)). a-b(iii) show the recovered coincidence rate counts, after varying the Gaussian pump waist (Fig. \ref{fig:lg1}.a(iv)).}}
 \label{fig:suppTolerance}
\end{figure*}

\textcolor{black}{We now take into account a case of crystal imperfection in order to assess the tolerance of the designed crystal under fabrication errors. To do this, we first let our algorithm find the optimal spatial modes of the crystal volume hologram for generating the quantum correlations of the desired quantum state with a fixed pump. After the learning phase, we deliberately add errors to the crystal structure (which mimics  crystal fabrication imperfections) and examine how does the desired quantum state is affected. We consider adding errors to the crystal coefficients in two ways (based on Eq. \ref{eq:parameterized}):
\begin{enumerate}[label=(\alph*)]
    \item $\alpha_\chi^n = \alpha_\chi^n (1. + \Delta_\sigma)$
    \item $\alpha_\chi^n = \alpha_\chi^n + \Delta_\sigma$
\end{enumerate}
We assume that the errors are normally distributed, i.e. $\Delta_\sigma\sim \mathcal{N}(0,\sigma^{2})$. In the first approach, there is a relative effect of the error on the amplitude of the coefficients. Although, the coefficients will always remain in the same subspace of the basis functions. In the second approach, we are no longer limited to the original subspace, but the additive noise is not correlated with the amplitude of the coefficients anymore. We present the results on the optimal quantum volume holograms, embedded in 3D NLPCs with a constant Gaussian pump beam, for generating the desired coincidence rate counts of maximally-entangled two-photon qubit $\ket{\psi}=(\ket{1,-1}+\exp(i\phi)\ket{-1,1})/\sqrt{2}$ quantum state, which was presented in the main text (section 3.2, Fig. \ref{fig:lg1}.a). Figs. \ref{fig:suppTolerance}a-b(i) present the imperfect 3D volume crystal hologram design of the original design (Fig. \ref{fig:lg1}.a(v)), for the two discussed approaches. Noise was added to the coefficients until the coincidence rate counts of maximally-entangled two-photon qubit were significantly impaired relative to the original design (Fig. \ref{fig:lg1}.a(ii)), as can be seen in Figs. \ref{fig:suppTolerance}a-b(ii). At this point, we maintained the imperfect 3D volume crystal hologram structure and tested if we can nearly recover the original system results, by modifying the Gaussian pump only (Fig. \ref{fig:lg1}.a(iv)). As can be seen in \ref{fig:suppTolerance}a-b(iii), the pump waist optimization nicely overcomes the fabrication errors, indicating the tolerance of the formed crystal. In other words, the fabrication errors diverge the model from the optimal minimum for generating the desired quantum state, but since the model was in global minimum rather than local minimum a slight variation in a different parameter of the system (pump waist) diverges the system back.}

\end{document}